\documentclass[twocolumn,dvipdfm]{pasj00}
\draft
\newcommand{{\ka}}{K$\alpha$}

\begin{document}
\SetRunningHead{Ishida et al.}{Suzaku Observations of SS Cyg}
\Received{2008/05/01}%{yyyy/mm/dd}
\Accepted{2008/05/02}%{yyyy/mm/dd}

\title{The Suzaku Observations of SS Cygni in Quiescence and Outburst}

%%% begin:list of authors
% Do NOT capitalize all letters in "textsc".
\author{Manabu \textsc{Ishida},\altaffilmark{1}
 Shunsaku \textsc{Okada},\altaffilmark{1,2}
 Takayuki \textsc{Hayashi},\altaffilmark{1,3}
 Ryoko \textsc{Nakamura},\altaffilmark{1,4} \\
 Yukikatsu \textsc{Terada},\altaffilmark{5}
 Koji \textsc{Mukai},\altaffilmark{6} and
 Kenji \textsc{Hamaguchi}\altaffilmark{6,7}
%  \thanks{Example: Present Address is xxxxxxxxxx}
}
\altaffiltext{1}{Institute of Space and Astronautical Science/JAXA, 3-1-1
Yoshinodai, Sagamihara, Kanagawa 229-8510}
\email{ishida@astro.isas.jaxa.jp}
\altaffiltext{2}{Space System Division, NEC Corporation,
 1-10 Nisshin-cho, Fuchu, Tokyo 183-8501}
\altaffiltext{3}{Department of Physics, Tokyo Metropolitan University,
 1-1 Minami-Osawa, Hachioji, Tokyo 192-0397}
\altaffiltext{4}{Department of Physics, Tokyo Institute of Technology,
 2-12-1 O-okayama, Megro-ku, Tokyo 152-8551}
\altaffiltext{5}{Department of Physics, Saitama University,
 255 Shimo-okubo, Sakura-ku, Saitama 338-8570}
\altaffiltext{6}{CRESST and X-ray Astrophysics Laboratory, NASA/GSFC,
 Greenbelt, MD 20771, USA}
\altaffiltext{7}{Department of Physics, University of Maryland, Baltimore County, 1000 Hilltop Circle, Baltimore, MD 21250}

%% `\KeyWords{}' always has to be placed before `\maketitle'.
 %Do NOT move this preamble from here!

\maketitle

\begin{abstract}
 We present results from the Suzaku observations of the dwarf nova
 SS~Cyg in quiescence and outburst in 2005 November. High sensitivity of
 the HXD PIN and high spectral resolution of the XIS enable us to
 determine plasma parameters with unprecedented precision. The maximum
 temperature of the plasma in quiescence
 $20.4^{+4.0}_{-2.6}\,\mbox{(stat.)}\pm 3.0\,\mbox{(sys.)}$~keV is
 significantly higher than that in outburst $6.0^{+0.2}_{-1.3}$~keV. The
 elemental abundances are close to the solar ones for the medium-Z
 elements (Si, S, Ar) whereas they decline both in lighter and heavier
 elements, except for that of carbon which is 2 solar at least. The
 solid angle of the reflector subtending over an optically thin thermal
 plasma is $\Omega^{\rm Q}/2\pi = 1.7\pm 0.2\,\mbox{(stat.)}\pm
 0.1\,\mbox{(sys.)}$ in quiescence. A 6.4~keV iron {\ka} line is
 resolved into narrow and broad components. These facts indicate that
 both the white dwarf and the accretion disk contribute to the
 reflection. We consider the standard optically thin boundary layer as
 the most plausible picture for the plasma configuration in
 quiescence. The solid angle of the reflector in outburst $\Omega^{\rm
 O}/2\pi = 0.9^{+0.5}_{-0.4}$ and a broad 6.4~keV iron line indicate
 that the reflection in outburst originates from the accretion disk and
 an equatorial accretion belt. The broad 6.4~keV line suggests that the
 optically thin thermal plasma is distributed on the accretion disk like
 solar coronae.
\end{abstract}

\section{Introduction}

Dwarf novae (DNe) are non-magnetic cataclysmic variables (CVs; binaries
between a white dwarf primary and a mass-donating late-type star) which
show optical outbursts typically with $\Delta m_V =$ 2--5 lasting
2--20~d with intervals of $\sim$10~d to tens of years
\citep{1995CAS....28.....W}. These outbursts can be explained as a
result of a sudden increase of mass-transfer rate within the accretion
disk surrounding the white dwarf due to a thermal-viscous instability
\citep{1974PASJ...26..429O,1981A&A...104L..10M,1982MNRAS.199..267B,1984PASP...96....5S,1993ApJ...419..318C,1996PASP..108...39O}.
A boundary layer (hereafter abbreviated as BL) is formed between the
inner edge of the accretion disk and the white dwarf where matter
transferred through the disk releases its Keplerian motion energy and
settles onto the white dwarf. BL is a target of EUV and X-ray
observations since its temperature becomes $T\simeq 10^5$--10$^8$~K.
\citet{1979MNRAS.187..777P} and \citet{1985ApJ...292..535P} have
predicted that radiation from the BL starts to shift from hard X-ray to
EVU when the high $\dot{M}$ front arrives at the inner edge of the disk,
because BL becomes optically thick to its own radiation. This prediction
has been verified by a number of multi-waveband coordinated observations
\citep{1979MNRAS.186..233R,1992MNRAS.257..633J,2003MNRAS.345...49W}.

The region around the inner edge of the disk is filled with a lot of
intriguing but still unresolved issues. Within the framework of the
standard accretion disk \citep{1973A&A....24..337S}, half of the
gravitational energy is released in the accretion disk, and hence, the
other half is released in BL. The observations in extreme-ultraviolet
band of VW~Hyi and SS~Cyg, however, revealed that the fractional energy
radiated from BL is only $<10$\% of the disk luminosity
\citep{1991ApJ...372..659M,1995ApJ...446..842M}.  According to the
classical theory, the temperature of BL in outburst is predicted to be
2--5$\times 10^5$~K \citep{1979MNRAS.187..777P}, whereas the temperature
estimated by ultraviolet and optical emission lines is constrained to a
significantly lower range 5--10$\times 10^4$~K
\citep{1991MNRAS.249..452H}. These discrepancies may be resolved if we
assume that BL is terminated not on the static white dwarf surface but
on a rapidly rotating accretion belt on the equatorial surface of the
white dwarf \citep{1978necb.conf...89P,1978A&A....63..265K}. Suggestions
of the accretion belt, rotating at a speed close to the local Keplerian
velocity, have been reported from a few DNe in outburst
\citep{1993ApJ...405..327L,1996ApJ...458..355H,1996ApJ...471L..41S,1997AJ....114.1165C,1998ApJ...497..928S}.
Mechanism has not been understood yet to drive a dwarf nova oscillation
(DNO) which is a highly coherent oscillation of soft X-ray and optical
intensities with a period of 3--40~s
\citep{1978ApJ...219..168R,1980ApJ...235..163C,1984ApJ...278..739C,1986A&A...158..233S,1998MNRAS.299..921M,1998PASP..110..403P,2001ApJ...562..508M}.
\citet{2002MNRAS.335...84W} try to understand DNO by assuming a
magnetically driven accretion onto the accretion belt by enhancing a
magnetic field the belt through a dynamo mechanism.

One of the unresolved outstanding issues may be the origin of a hard
X-ray optically thin thermal emission in outburst, since BL is believed
to be optically thick. In order to identify its emission site, and to
obtain some new insight on BL in quiescence as well, we planned to
observe SS~Cyg both in quiescence and outburst with the X-ray
observatory Suzaku \citep{2007PASJ...59S...1M}. SS~Cyg is a dwarf nova
in which the $1.19\pm 0.02\MO$ white dwarf and the $0.704\pm 0.002 \MO$
secondary star \citep{1990MNRAS.246..654F} are revolving in an orbit of
$i = 37^\circ\pm 5^\circ$ \citep{1983PhDT........14S} with a period of
6.6 hours. The distance to SS~Cyg is measured to be $166\pm12$~pc using
HST/FGS parallax \citep{1999ApJ...515L..93H}. SS~Cyg shows an optical
outburst roughly every 50 days, in which $m_V$ changes from 12th to 8th
magnitude. The optically thin to thick transition of BL has clearly been
detected with coordinated observations of optical, EUVE, and RXTE
\citep{2003MNRAS.345...49W}. In \S~2, we show an observation log and
procedure of data reduction. In \S~3, detail of our spectral analysis is
explained. Owing to a high spectral resolution of the X-ray Imaging
Spectrometer (XIS; \cite{2007PASJ...59S..23K}) and a high sensitivity of
the Hard X-ray Detector (HXD; \cite{2007PASJ...59S..35T};
\cite{2007PASJ...59S..53K}) over 10~keV enable us to determine spectral
parameters of hard X-ray emission of SS~Cyg with unprecedented
precision. In \S~4, we discuss on the emission site and its spatial
extension both in quiescence and outburst by utilizing the spectral
parameters, a 6.4~keV iron line parameters in particular. We summarize
our results and discussions in \S~5.

\section{Observation and data reduction}

\subsection{Observations}

\begin{figure}[htb]
\begin{center}
 \includegraphics[width=0.6\textwidth,bb=0 0 1012 783]{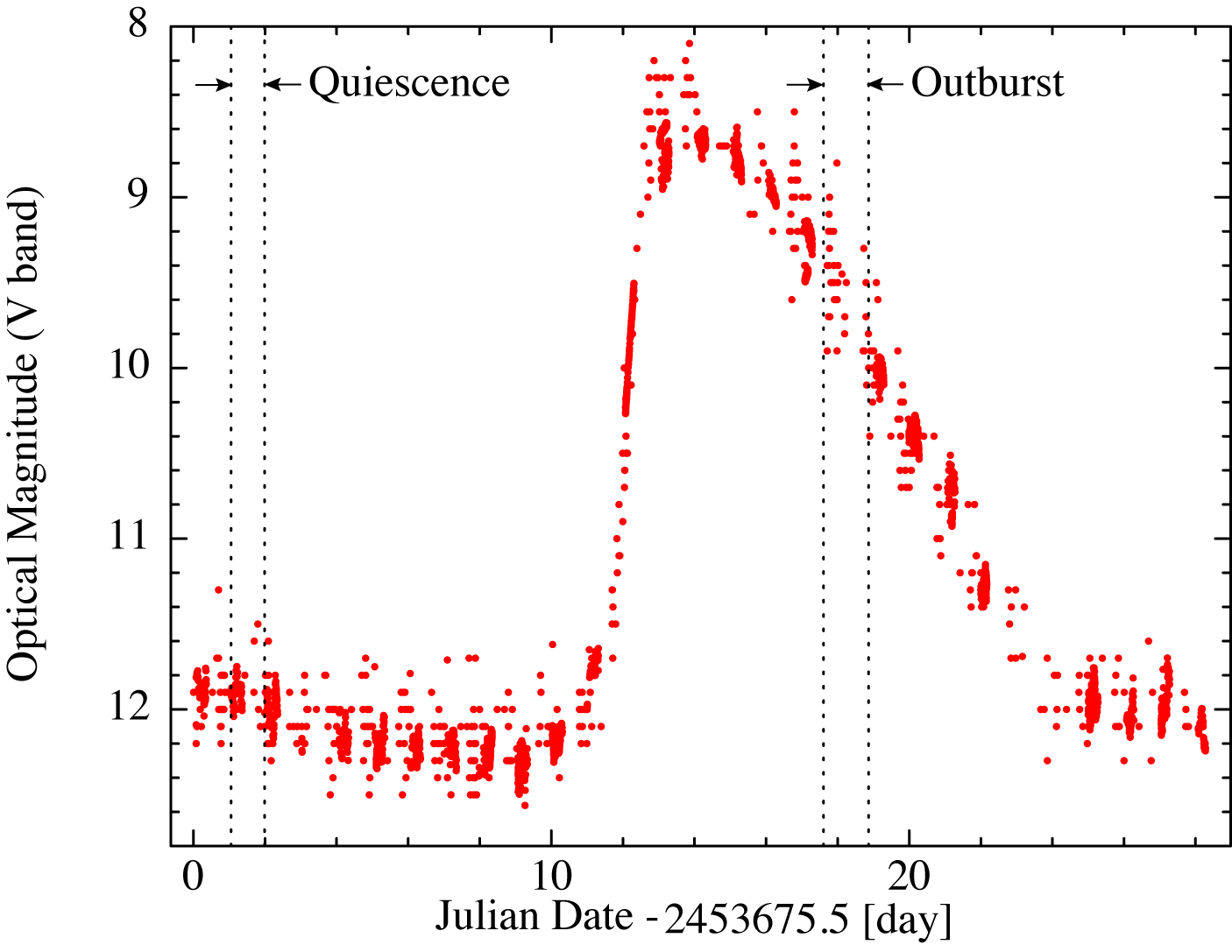}
\end{center}
\caption{A V-band light curve of SS Cyg covering the Suzaku
 observations.  The observation periods of quiescence and outburst are
 segmented with dashed vertical lines. Taken from the AAVSO home page.
 \label{fig:optlc}}
\end{figure}

The Suzaku observations of SS~Cygni in quiescence and outburst were
carried out during 2005 November 2 01:02--23:39(UT), and 2005 November
18 14:15(UT)--November 19 20:45(UT), respectively, as part of the
performance verification programme \citep{2007PThPS.169..178I}.  In
Fig.~\ref{fig:optlc}, we show an optical light curve covering our SS~Cyg
observations taken from the home page of American Association of
Variable Star Observers (AAVSO)\footnote{http://www.aavso.org/}.  The
outburst observation was performed $\sim$two days after the optical
maximum.  The observation log is summarized in table~\ref{tab:obslog}.
%
% Observation Log
\begin{table}[htb]
\setlength{\tabcolsep}{3pt}
\caption{Suzaku observation log of SS Cyg \label{tab:obslog}}
\begin{center}
\begin{tabular}{llllllll}
\hline\hline
State & Seq. \# & Observation Date & Pointing & Detector\footnotemark[$\#$]
 & Mode & Exp.\footnotemark[$\ast$]
 & Intensity\footnotemark[$\dagger$] \\
      &    & [UT] & & & [ks] & [c s$^{-1}$] \\ \hline
Quiescence & 40006010
 & 2005 Nov. 02 01:02 --- 02 23:39 & XIS nom. & FI & Normal & 39.3 & 3.308$\pm$0.005\\
 & & & & BI  & Normal & 39.3 & 4.480$\pm$0.011\\
 & & & & PIN & Normal & 27.0 & 0.163$\pm$0.005\\
Outburst   & 40007010
 & 2005 Nov. 18 14:15 --- 19 20:45 & XIS nom. & FI & Normal & 56.0 & 1.580$\pm$0.003\\
 & & & & BI  & Normal & 56.0 & 2.427$\pm$0.007\\
 & & & & PIN & Normal & 47.9 & 0.028$\pm$0.003\\
\hline
\multicolumn{8}{@{}l@{}}{\hbox to 0pt{\parbox{180mm}{\footnotesize
\footnotemark[$\#$] FI: Frontside-Illuminated CCD (XIS0, 2, 3), BI:
 Backside-Illuminated CCD (XIS1), PIN: HXD PIN detector.
\par\noindent
\footnotemark[$\ast$] Exposure time after data screening.
\par\noindent
\footnotemark[$\dagger$] Intensity in the 0.4--10~keV (FI and BI CCDs)
 and 12--40~keV bands (PIN) with a 1$\sigma$ statistical errors. The
 systematic error of the PIN background (5\% of the NXB) is
 0.024~c~s$^{-1}$ in this band.
}\hss}}
\end{tabular}
\end{center}
\end{table}
Throughout the observations, XIS was operated in the normal 5$\times$5
and 3$\times$3 editing modes during the data rate SH/H and M/L,
respectively, with no window/burst options, while the HXD PIN was
operated with the bias voltage of 500~V for all the 64 modules. We do
not use the HXD GSO data, because we did not receive any significant
signal from SS~Cyg throughout. The source was placed at the XIS nominal
position in both observations where the observation efficiency of all
the XRTs are more than 97\% of maximum throughput
\citep{2007PASJ...59S...9S}. As already known from previous
observations (e.g.\cite{2003MNRAS.345...49W}) the hard X-ray flux was
smaller in outburst than in quiescence.
% Source of SS Cyg is located at ($\alpha$, $\delta$)=($325^{\circ}.6780$,
% $43^{\circ}.5860$) in J2000 coordinates. 

\subsection{Data screening
\label{sec:datascreening}}

In data reduction, we use event files that are created with the pipeline
processing software (revision 1.2.2.3), and the analysis software
package HEASOFT (version 6.1.3). For the XIS, we selected the events
with grades 0, 2, 3, 4, and 6. We do not use bad pixels or columns where
charge transfer efficiency is too low to detect/transfer X-ray
signals. We also discarded the data taken during time intervals of
maneuvers, low data rate, passage of SAA (South Atlantic Anomaly), the
field of view being occulted by the earth or watching bright earth rim,
and the pointing position being away from the source by more than
$\timeform{1'.5}$. Finally, the cleaned event files are created by
removing hot/flickering pixels. In addition to the criteria described
above, we accept the data without telemetry saturation. After these data
selection, we combined the data of 3$\times$3 and 5$\times$5 editing
modes before starting analysis. Resultant exposure time is 39.3~ks and
56.0~ks in quiescence and outburst, respectively. For creating a
cleaned event file for the HXD PIN, on the other hand, we further
removed the time interval during cut-off rigidity is less than
6~GeV~c$^{-1}$.

In extracting photons from the source with the XIS, we take a circular
integration region with a radius of 250 pixels ($\timeform{4.'34}$)
centered at the source, which includes more than 96 \% of the X-ray
events, while an annulus with an outer radius of 432 pixels
($\timeform{7.'50}$) around the source region is adopted for the
background-integration region. In view of statistics, we took the area
of the background-integration region twice as large as that of the
source-integration region. For the HXD-PIN, on the other hand, we
accumulate a non-X-ray background (NXB) spectrum from a simulated NXB
event file\footnote{ftp://ftp.darts.jaxa.jp/pub/suzaku/ver1.2/}
published by the HXD team. In addition to the NXB, we need to consider
the cosmic X-ray background (CXB), for which we adopt the empirical
model spectrum constructed on the basis of the {\it HEAO} observations
\citep{1987PhR...146..215B},
\begin{eqnarray}
 f_{\rm CXB}(E) &=& 
  9.0\times \left(\frac{E}{3~{\rm keV}}\right)^{-0.29}\times
 \exp\left( -\frac{E}{40~{\rm keV}}\right)
 \quad{\rm erg~cm^{-2}~str^{-1}~keV^{-1}},
\end{eqnarray}
and create a CXB spectrum from this model by the {\tt fakeit} command in
the spectral fit software XSPEC \citep{1996ASPC..101...17A} with an
exposure time of 1~Ms. In doing this, we adopt the PIN flat sky
response. The background spectrum for the PIN is created by combining
the NXB and CXB spectra using the {\tt mathpha} command in the FTOOLS
package. Since the CXB level is $\sim$5\% of the NXB level, we ignore
sky-to-sky variation of the CXB. The counting rates listed in
table~\ref{tab:obslog} are those after all the data screening described
in this section are applied.

\subsection{Light Curves 
\label{sec:lightcurve}}

\begin{figure}[htb]
\centerline{
 \includegraphics[width=0.49\textwidth,bb=0 0 978 835]{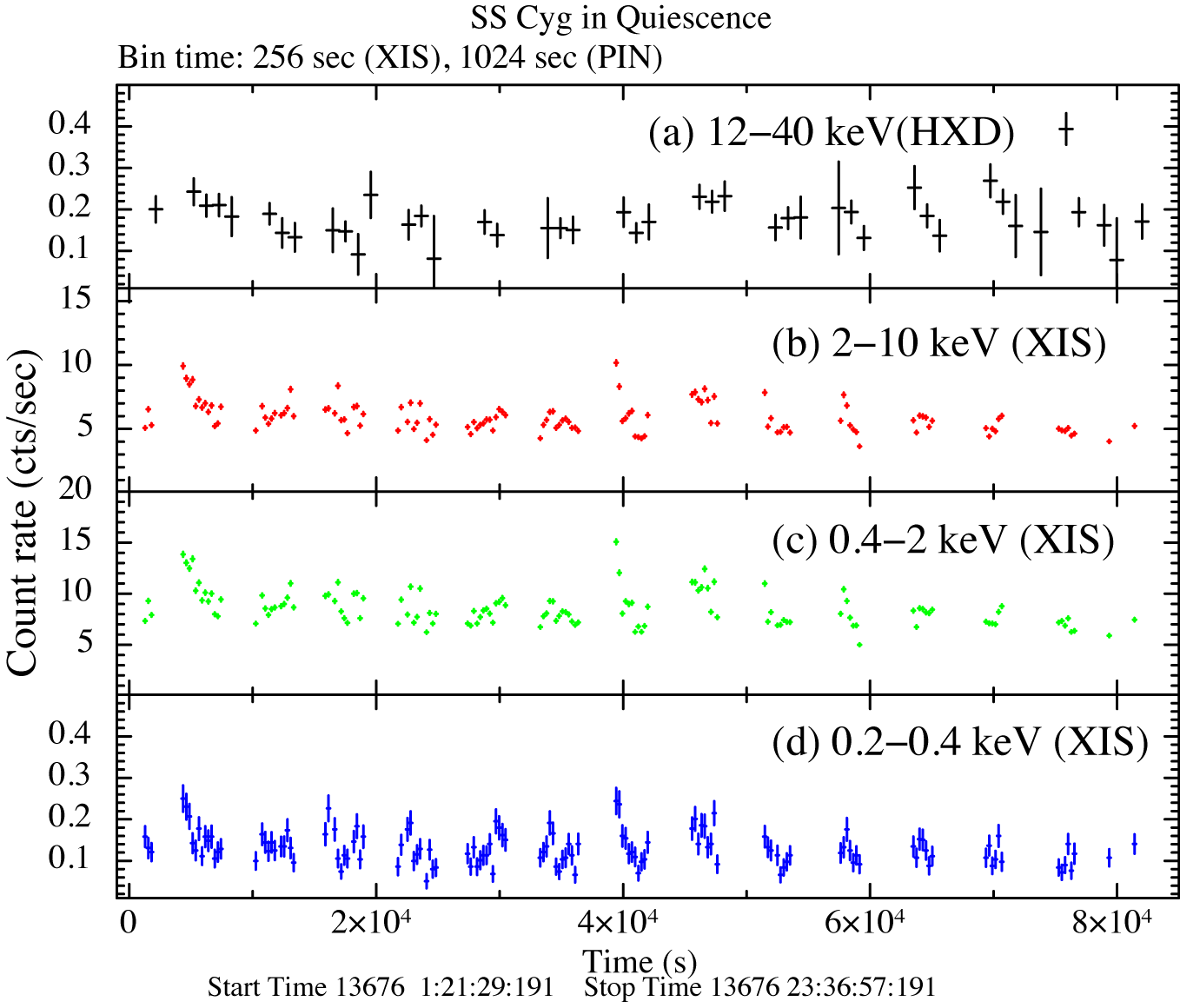}
 \includegraphics[width=0.49\textwidth,bb=0 0 978 835]{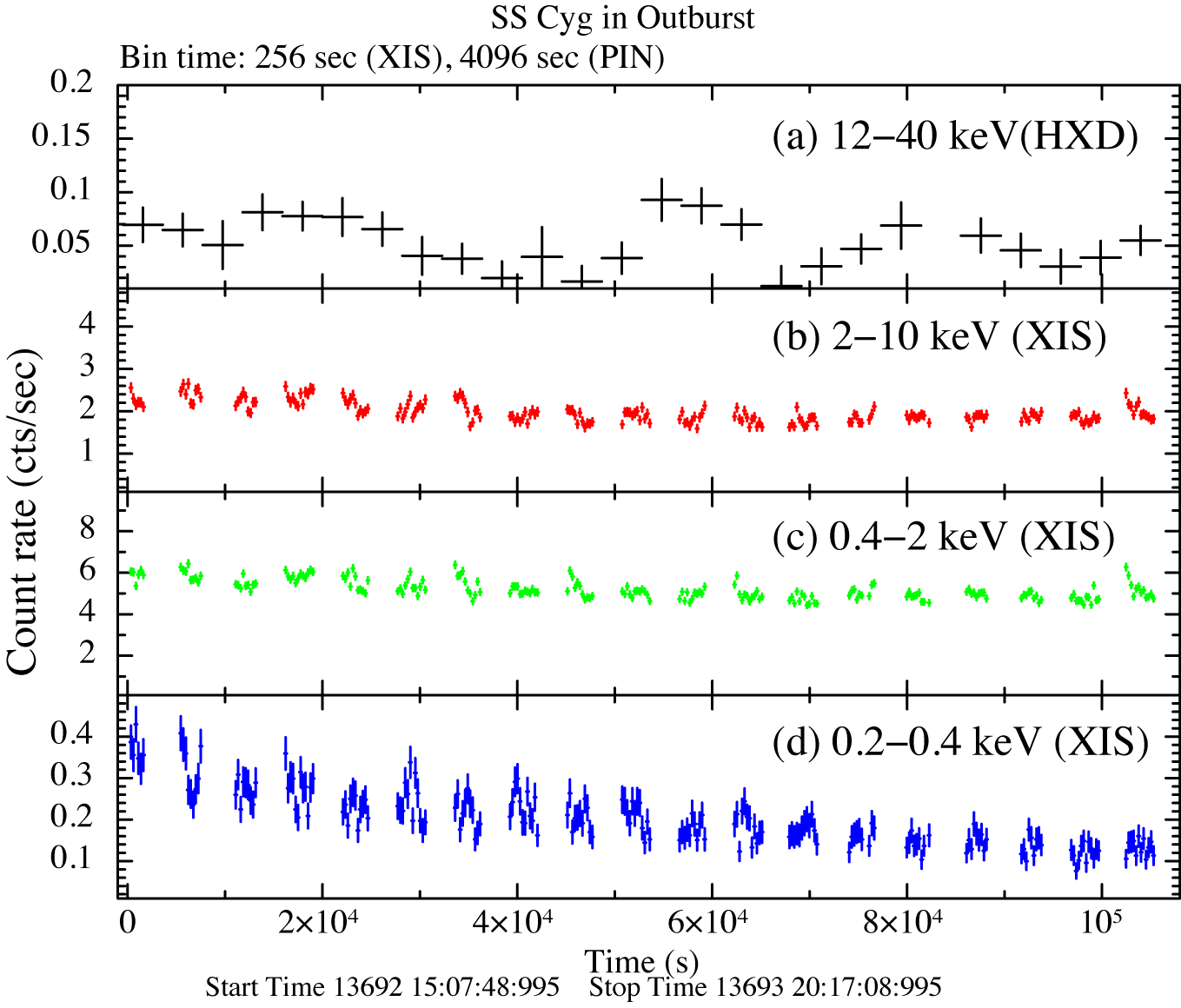}
 }
\caption{Background-subtracted light curves of SS Cyg in quiescence
 (left) and outburst (right) in the bands (a) 12--40~keV, (b) 2--10~keV,
 (c) 0.4--2~keV, and (d) 0.2--0.4~keV. Data from all the XIS modules
 are combined to make the light curves in the panels (b) through
 (d). The bin size of the XIS is 256~sec both in quiescence and outburst
 whereas that of the HXD PIN is 1024~sec in quiescence and 4096~sec in
 outburst. The origin of time is MJD 57676 01:23:49 and MJD 57692 14:56:10,
 respectively. \label{fig:lc}}
\end{figure}

We show energy-resolved and background-subtracted light curves of SS~Cyg
in quiescence and outburst in Fig.~\ref{fig:lc}.
The source and background photons are accumulated using the integration
regions explained in \S~\ref{sec:datascreening}.  The data from the four
XIS modules are combined to generate the light curves in the three
energy bands 0.2--0.4~keV, 0.4--2.0~keV, and 2.0--10.0~keV. The light
curve of the HXD (PIN) is created in the energy range 12--40~keV where
the sensitivity of HXD becomes the maximum. The mean counting rates are
summarized in table~\ref{tab:obslog}. In outburst, the average intensity in the
12-40~keV band is 5.9\% of the NXB intensity of the PIN, whereas there
remains 5\% systematic error in the NXB model intensity. Hence, the
detection of the PIN in outburst is marginal. In the energy bands above
0.4~keV, the source is brighter in quiescence than in outburst, as
demonstrated by \citet{2003MNRAS.345...49W}, and is more variable. The
flux below 0.4~keV is, on the other hand, higher in outburst. Moreover,
the source obviously declines throughout the observation, whereas the
intensity is nearly constant in the higher energy bands. These facts
suggest that Suzaku detected high energy end of the emission from the
optically thick BL which appears in outburst
\citep{1977AcA....27..235T,1977MNRAS.178..195P,1979MNRAS.187..777P,1985ApJ...292..535P,2003MNRAS.345...49W}.

\subsection{Average spectra
\label{sec:spectra}}

\begin{figure}[htb]
\centerline{
 \includegraphics[width=0.49\textwidth,bb=0 0 1040 781]{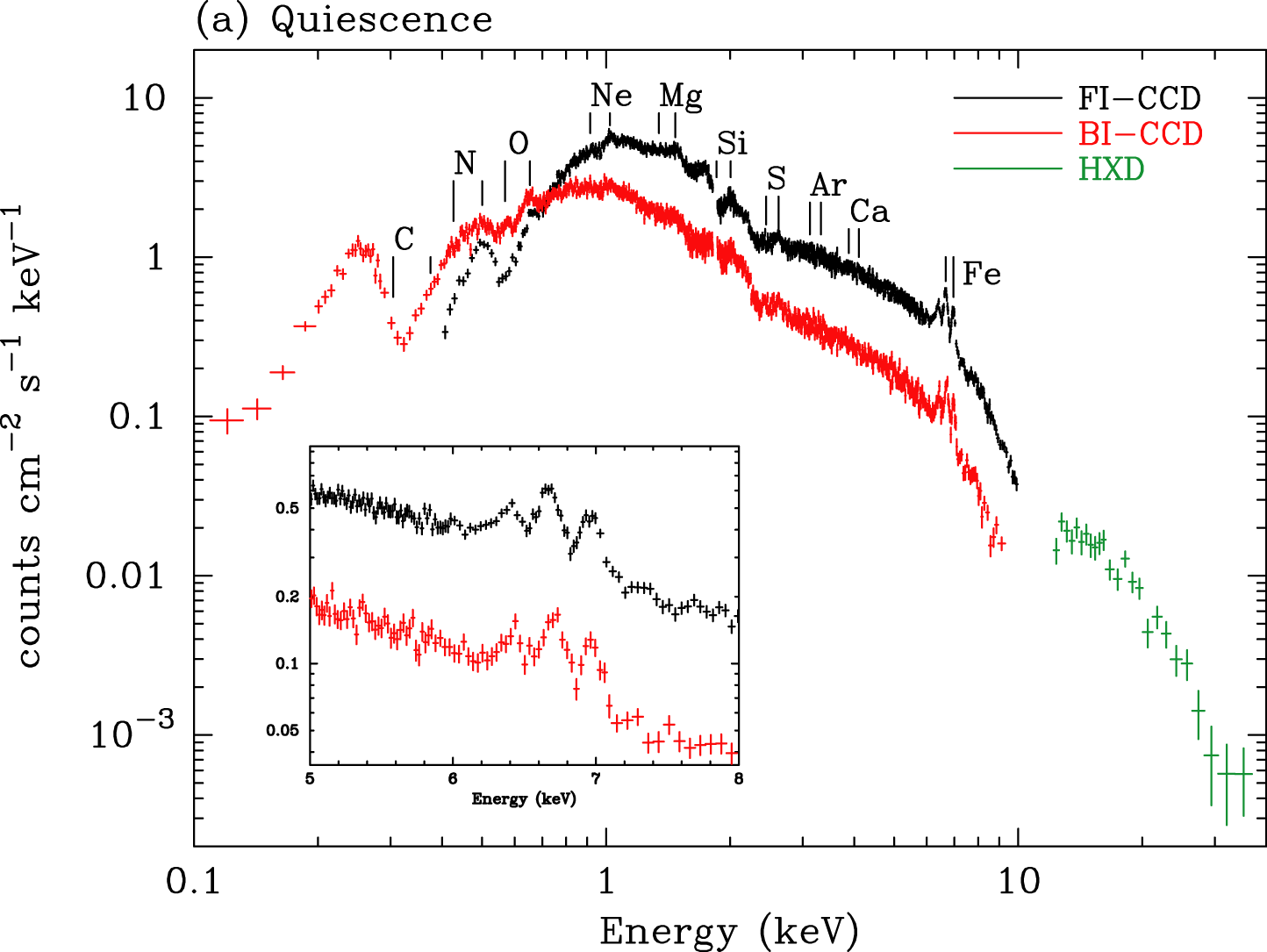}
 \includegraphics[width=0.49\textwidth,bb=0 0 1040 781]{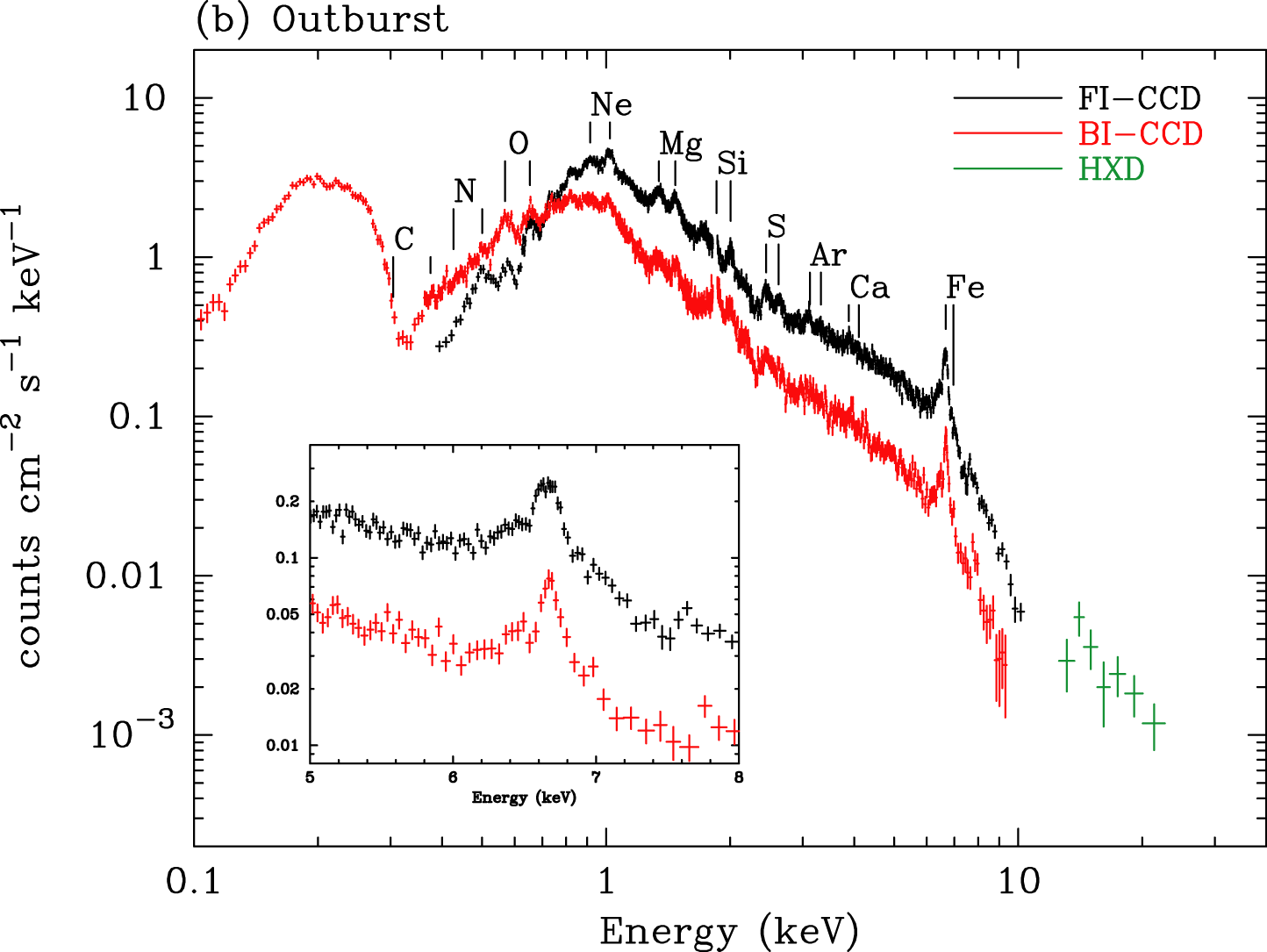}
 } 
\caption{Averaged spectra of SS~Cyg in quiescence and outburst. The
 source and background integration regions are explained in
 \S~\ref{sec:datascreening}. The data from XIS0, 2, and 3 are combined
 into a single FI-CCD spectrum whereas BI-CCD spectrum originates from
 XIS1. The insets are the blowups in the 5-8~keV band showing profiles
 of iron {\ka} emission lines.
\label{fig:avespec}}
\end{figure}

In Fig.~\ref{fig:avespec}, we show averaged spectra of SS~Cyg from the
XIS and the HXD-PIN in quiescence and outburst.
The source and background photons are accumulated using the integration
regions explained in \S~\ref{sec:datascreening}. The data from XIS0, 2,
and 3 are combined into a single FI-CCD spectrum. The BI-CCD spectrum
originates solely from the XIS1 data.

SS~Cyg is detected at least up to $\sim$30~keV with the HXD-PIN in
quiescence, whereas the detection of the PIN in outburst is marginal as
noted in \S~\ref{sec:lightcurve}. From the inset, iron {\ka} emission
lines are clearly resolved into 6.4, 6.7, and 7.0~keV components. The
6.4~keV line indicates reflection of the hard X-ray emission from the
white dwarf and possibly from the accretion disk, as pointed out by
\citet{1997MNRAS.288..649D}. Note that the 6.4~keV line is broad in
outburst. Except for the iron emission lines, only weak signs of H-like
{\ka} lines are visible from the other elements in quiescence. The
outburst spectra, on the other hand, are softer than those in
quiescence, and are characterized by H-like and He-like {\ka} emission
lines from nitrogen to iron \citep{2008ApJ...680..695O}.  From the
inset, the He-like iron {\ka} line at 6.7~keV is much stronger than the
H-like line at 7.0~keV, in contrast to the quiescence spectra. These
facts indicate that the plasma has a temperature distribution, and that
the average plasma temperature is significantly lower in outburst than
in quiescence. The decaying soft component in the light curves
(Fig.~\ref{fig:lc}) appears as an excess soft emission below
$\sim$0.3~keV in the BI-CCD spectrum in outburst.

\section{Analysis and Results}

\subsection{Spectral components and their models}

\begin{figure}[htb]
\centerline{
 \includegraphics[width=0.6\textwidth,bb=0 0 1080 810]{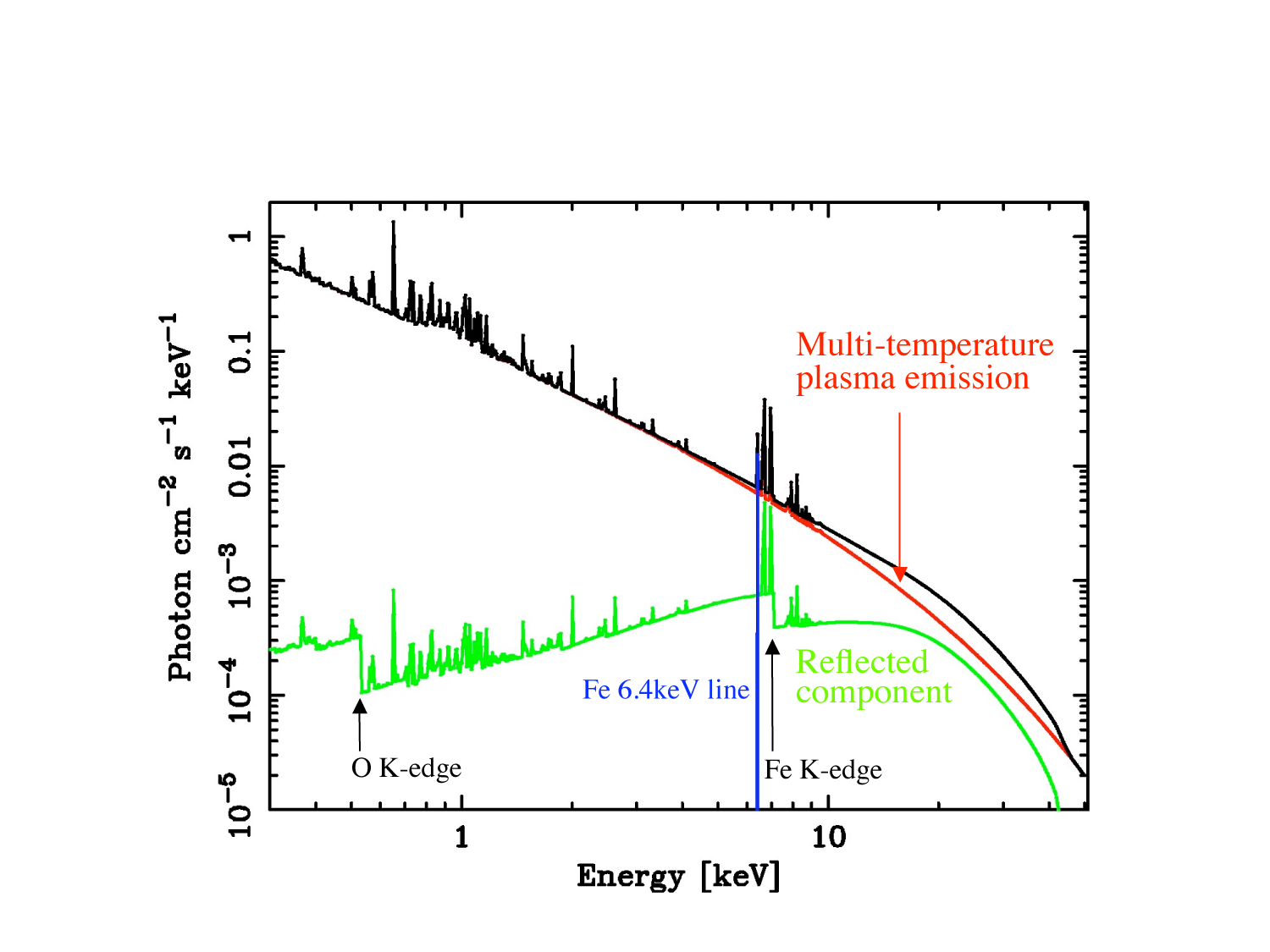}
}
\caption{Schematic view of the spectral component of SS~Cyg in the
 0.2--40~keV band. \label{fig:speccomp}}
\end{figure}

In Fig.~\ref{fig:speccomp}, we show schematic view of spectral
ingredients of SS~Cyg in the 0.2--40~keV band.
This picture is originally proposed by \citet{1997MNRAS.288..649D} on
the basis of their analysis of the Ginga and ASCA data of SS~Cyg. The
hard spectrum of SS Cyg is primarily composed of an optically thin
thermal plasma emission with a temperature distribution
(\S~\ref{sec:spectra}), for which \citet{1997MNRAS.288..649D} introduced
a power-law type differential emission measure (DEM) model as
\begin{equation}
 d({\rm EM}) \propto \left( \frac{T}{T_{\rm max}} \right)^\alpha\,d(\log T)
  \propto \left( \frac{T}{T_{\rm max}} \right)^{\alpha-1}\,dT,
\label{eq:dem}
\end{equation}
where $T_{\rm max}$ is the maximum temperature of the plasma. The model
they used, named {\tt cevmkl} in XSPEC, is an optically thin thermal
plasma emission model ({\tt mekal} in XSPEC,
\cite{1985A&AS...62..197M,1986A&AS...65..511M,1995ApJ...438L.115L,1996uxsa.coll..411K})
convoluted with this temperature distribution. Since it can successfully
represent $\sim$30 spectra of the dwarf novae
\citep{2005MNRAS.357..626B} observed with ASCA, we adopt this model as
well in this paper. In addition to this main component, its reflection
from the white dwarf (and the accretion disk) occupies significant
fraction of the observed X-ray flux, as required from detection of a
fluorescent iron {\ka} line at 6.4~keV. To represent the reflection, we
adopt the model {\tt reflect} \citep{1995MNRAS.273..837M}, which is a
convolution-type model describing reflectivity of neutral material. In
summary, the X-ray spectra of SS~Cyg are composed of the
multi-temperature optically thin thermal plasma component, its
reflection from the white dwarf (and the accretion disk), and the
fluorescent iron emission line at 6.4~keV.

\subsection{Evaluation of plasma parameters \label{sec:plasmaparam}}

Our purpose is to constrain the geometry of the optically thin thermal
plasma in SS~Cyg. In doing this, we can utilize the equivalent width
(EW) and the profile of the 6.4~keV emission line. The EW reflects a
covering fraction of a reflector viewed from the plasma
\citep{1986LNP...266..249M}. The profile contains information of motion
of the reflector. From continuum analysis, the covering fraction of the
reflector can independently be evaluated.

To utilize these methods, however, it is essential to know a priori the
iron abundance $Z_{\rm Fe}$. For its estimation, we generally make use
of the relative intensities and the EWs of the He-like and H-like iron
{\ka} emission lines at 6.7~keV and 7.0~keV, respectively. In deriving
$Z_{\rm Fe}$ from these quantities, we need to know the plasma emission
parameters such as $T_{\rm max}$ and $\alpha$ through spectral fitting
of the {\tt cevmkl} model to the observed data. One may suspect that it
is enough to fit the spectrum containing the He-like and H-like lines
locally with a single temperature optically thin thermal plasma
model. This does not work properly, however, since the temperature of
the plasma in SS~Cyg is distributed in such a wide range, from $T_{\rm
max}\simeq 20$~keV (see below) down to $\sim$0.1~keV indicated by the
nitrogen and oxygen lines, that plasma emission components whose
temperatures are too high enhance only the continuum level and dilute
the iron emission lines, thereby resulting in a lower abundance
estimation \citep{1997MNRAS.288..649D}.

\begin{figure}[htb]
\centerline{
 \includegraphics[width=0.33\textwidth,bb=0 0 976 747]{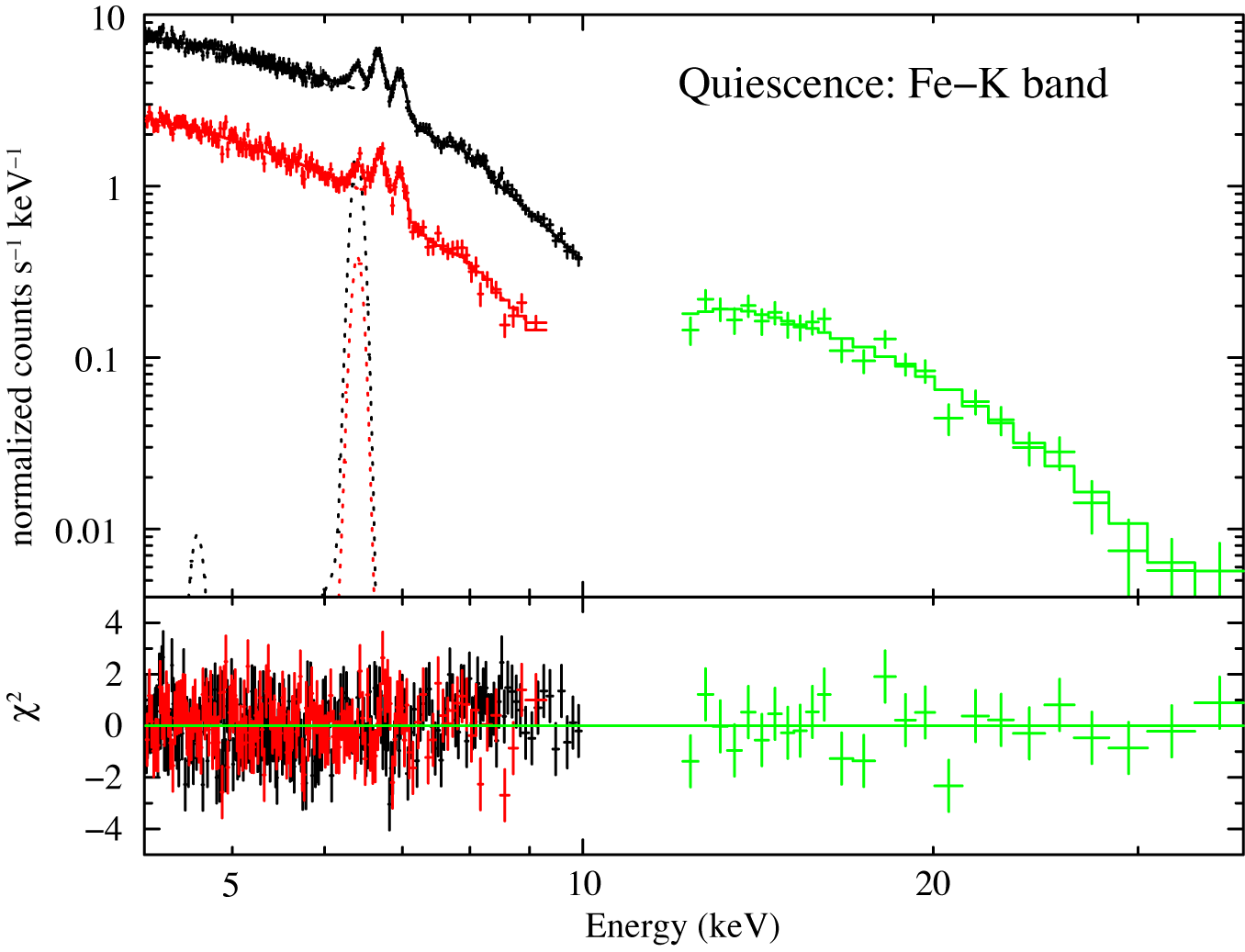}
 \includegraphics[width=0.33\textwidth,bb=0 0 976 737]{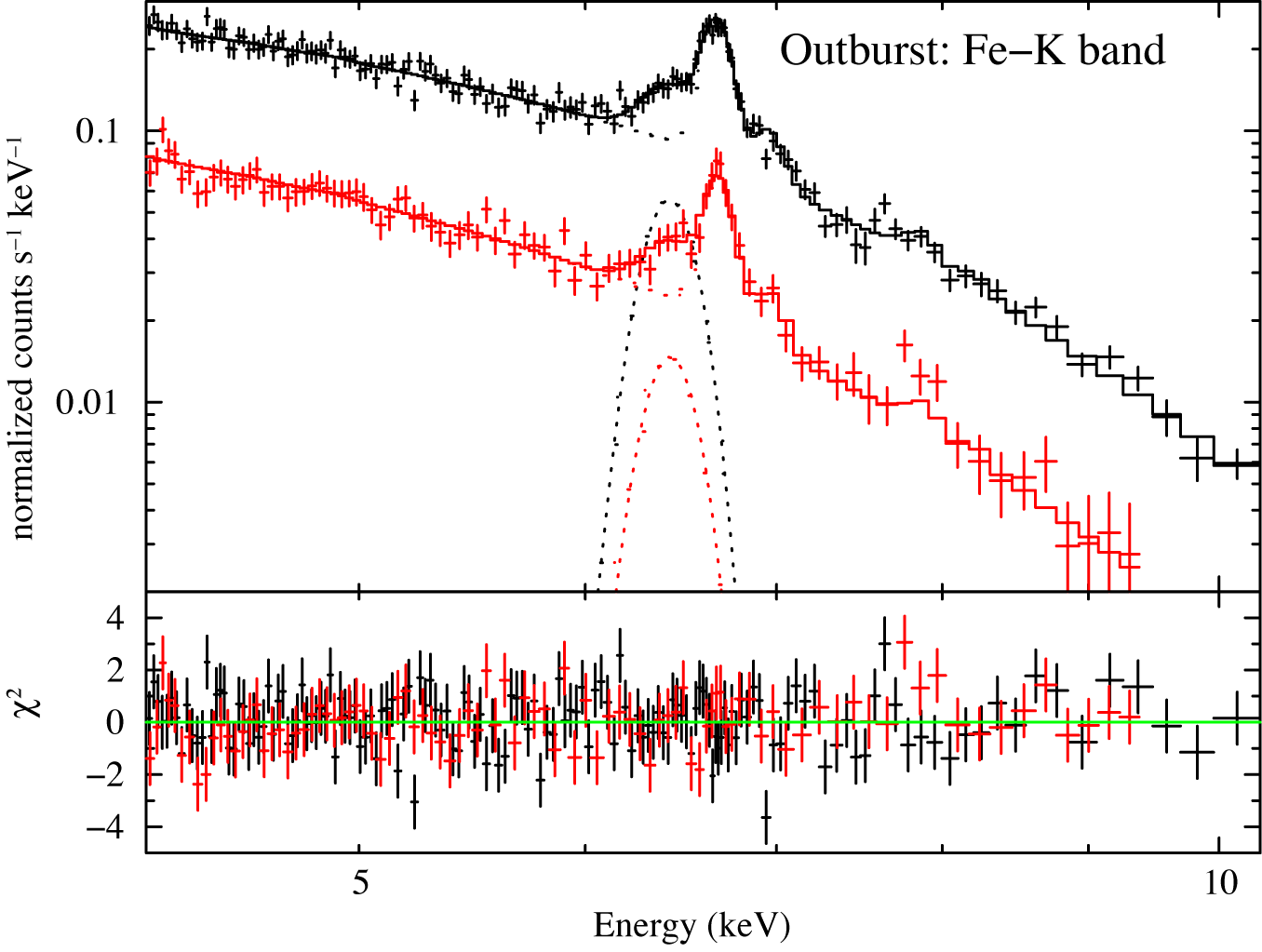}
 \includegraphics[width=0.33\textwidth,bb=0 0 976 737]{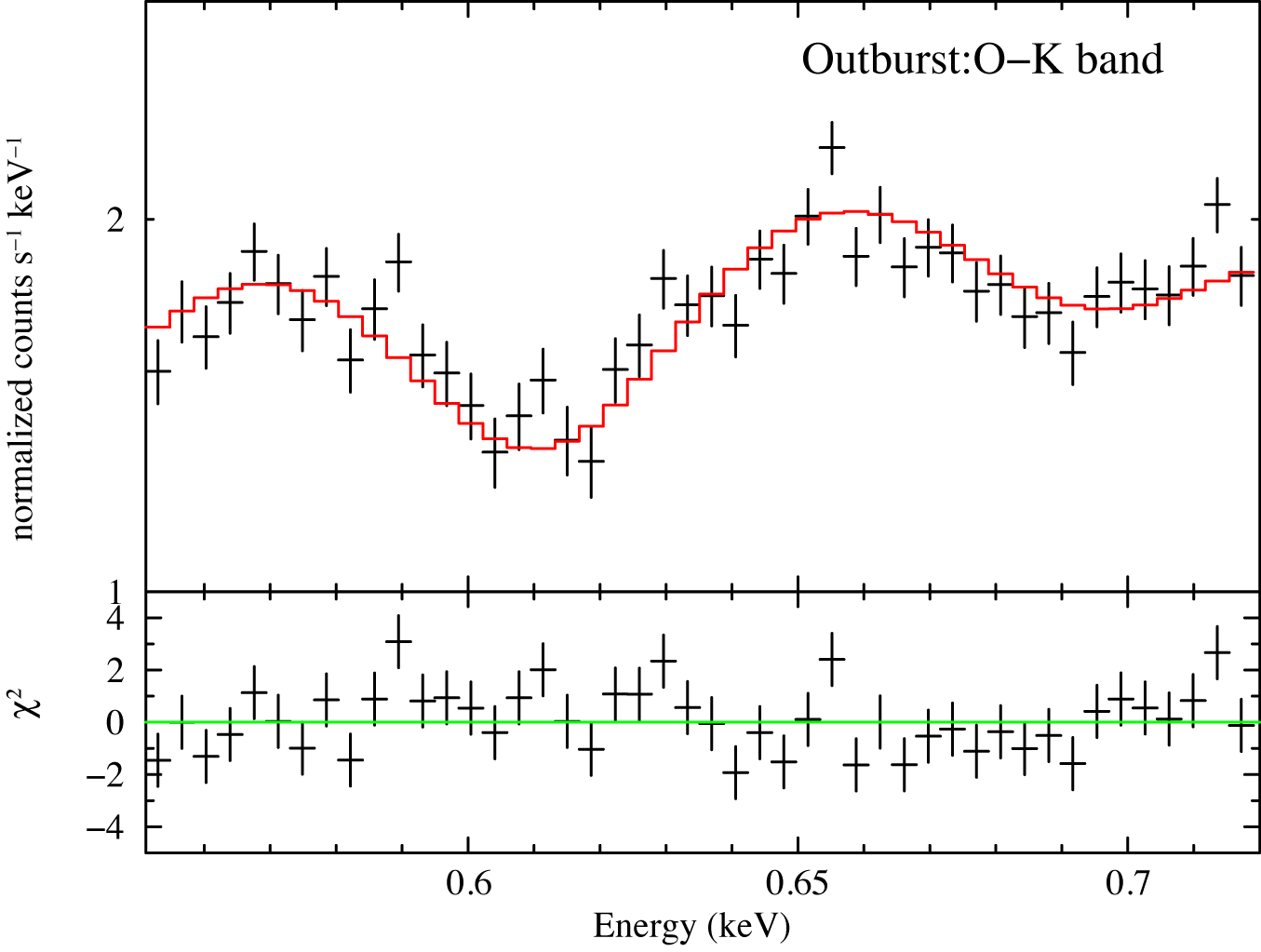}
 }
\caption{Result of the simultaneous fits of the {\tt cevmkl} plus
 reflection model to the 4.2--40~keV band in quiescence, the 4.2--10~keV
 and 0.55--0.72~keV bands in outburst. Black, red, and green colors are
 used for the data and the model of the FI-CCD, BI-CCD, and PIN,
 respectively, except for the outburst 0.55--0.72~keV spectrum where
 only the BI-CCD data are utilized.  \label{fig:simulfit}}
\end{figure}

\begin{figure}[htb]
\centerline{
 \includegraphics[width=0.6\textwidth,bb=0 0 1603 801]{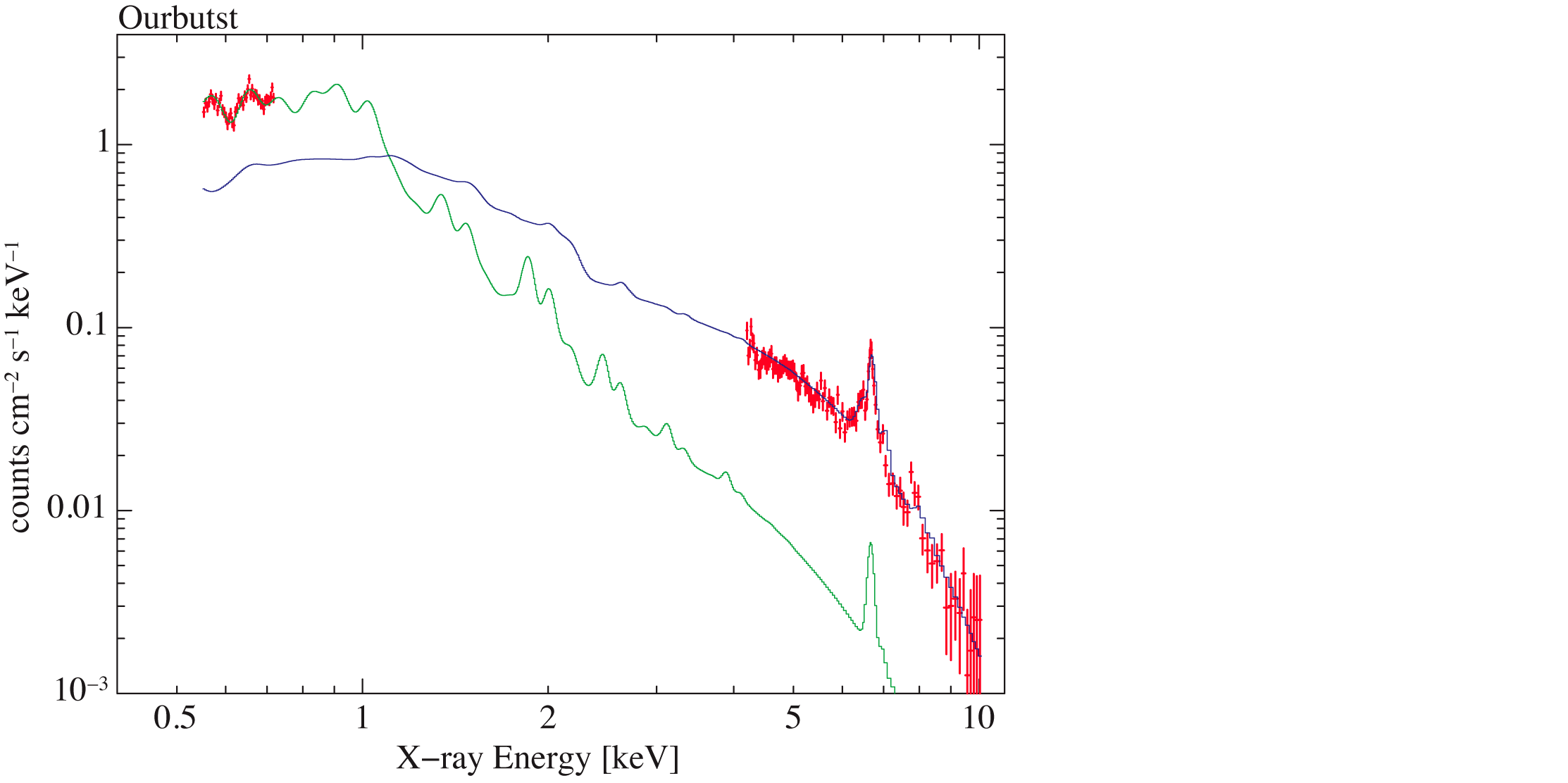}
}
\caption{The outburst spectra and the models used to determine the
 parameters of the plasma emission and the reflection. Only the data of
 the XIS1 are shown for clarity in red. The green and the blue curves
 show the best-fit models in the 0.55--0.72~keV and 4.2-40~keV bands,
 respectively, whose parameters are summarized in table~\ref{fig:simulfit}.
\label{fig:consistency}}
\end{figure}

To know the plasma emission parameters $T_{\rm max}$ and $\alpha$,
however, we need to know the metal abundances, the iron abundance
($Z_{\rm Fe}$) in particular and the oxygen abundance ($Z_{\rm O}$) as
well, and the covering fraction of the reflector ($\Omega$), because
they couple with each other in the spectrum above 7~keV. At this stage,
we have found that the plasma emission parameters ($T_{\rm max}$ and
$\alpha$) and the reflection parameters ($\Omega$, $Z_{\rm Fe}$ and
$Z_{\rm O}$) depend on each other, and need to determine them in a
self-consistent manner. Accordingly, we carried out a combined spectral
fit among the quiescence and outburst spectra in selected bands crucial
to determine $T_{\rm max}$, $\alpha$, $\Omega$, $Z_{\rm Fe}$, and
$Z_{\rm O}$. They are the quiescence spectra in the 4.2--40~keV band
(sensitive to $T^{\rm Q}_{\rm max}$, $\alpha^{\rm Q}$, $Z_{\rm Fe}$, and
$\Omega^{\rm Q}$), the outburst spectra in the 4.2--10~keV (sensitive to
$T^{\rm O}_{\rm max}$, $\alpha^{\rm O}$, $Z_{\rm Fe}$, and $\Omega^{\rm
O}$) and that in 0.55--0.72~keV bands (sensitive to $Z_{\rm O}$ and
$\alpha^{\rm O}$), where the superfixes `Q' and `O' indicate quiescence
and outburst, respectively. We ignore the intermediate 0.72--4.2~keV
band in outburst, because (1) this band is of no use to evaluate the
parameters listed above. Including this band may rather introduce
unexpected systematic errors in the parameters of interest, (2) this
band includes the emission lines from Ne to Ca whose abundances are
uncertain until we determine $T^{\rm O}_{\rm max}$ and $\alpha^{\rm O}$,
and (3) as will be found later in table~\ref{tab:simulfit} and
\ref{tab:abundances}, values of $\alpha^{\rm O}$ are different for
different elements. The entire outburst spectrum can not be fit with a
single $\alpha$ {\tt cevmkl} model.  We set $T_{\rm max}$, $\alpha$, and
$\Omega$ free to vary in quiescence and outburst separately, because the
geometry of the plasma, and hence its temperature distribution may be
different, whereas we set $Z_{\rm Fe}$ and $Z_{\rm O}$ common between
quiescence and outburst. Note that we do not use the PIN spectrum in
outburst, because the detection is marginal due to the NXB uncertainty
(\S~\ref{sec:lightcurve} and \ref{sec:spectra}). The result of the fits
is shown in Fig.~\ref{fig:simulfit}, and its best-fit parameters are
summarized in table~\ref{tab:simulfit}.
\begin{table}[htb]
\caption{Best-fit parameters of the simultaneous fit of the {\tt cevmkl}
 plus reflection model to the quiescence and outburst
 spectra.\label{tab:simulfit}}
\begin{center}
\begin{tabular}{lccc} \hline\hline
Phase               & Quiescence & \multicolumn{2}{c}{Outburst} \\ 
Energy band [keV]   & 4.2--40    & 4.2--10 & 0.55--0.72 \\ \hline
$T_{\rm max}$$^\ast$ [keV]  & $20.4^{+4.0}_{-2.6}\,(\pm 3.0)$ &
	 \multicolumn{2}{c}{$6.0^{+0.2}_{-1.3}$} \\
$\alpha$$^\dagger$       & $0.7^{+0.3}_{-0.1}\,(\pm 0.1)$ & $5.8^{+2.6}_{-1.4}$ &
	     $-0.36^{+0.02}_{-0.03}$ \\ 
$\Omega/2\pi$$^\ddagger$ & $1.7\pm 0.2\,(\pm 0.1)$ &
	 \multicolumn{2}{c}{$0.9^{+0.5}_{-0.4}$} \\
$Z_{\rm Fe}$$^\S$        & \multicolumn{3}{c}{$0.37^{+0.01}_{-0.03}\,(\pm
     0.01)$} \\
$Z_{\rm O}$$^\|$         & \multicolumn{3}{c}{$0.46^{+0.04}_{-0.03}\,(\pm
     0.01)$} \\
$\chi^2$ (d.o.f.)   & \multicolumn{3}{c}{728.2 (657)} \\
\hline
\multicolumn{4}{@{}l@{}}{\hbox to 0pt{\parbox{85mm}{\footnotesize
 Note --- Errors in parentheses are systematic errors associated with the
 NXB uncertainty of the HXD-PIN.
 \par\noindent
 \footnotemark[$*$] Maximum temperature of the optically thin thermal plasma.
 \par\noindent
 \footnotemark[$\dagger$] Power of DEM as $d({\rm EM}) \propto (T/T_{\rm
 max})^{\alpha -1}dT$.
 \par\noindent
 \footnotemark[$\ddagger$] Solid angle (covering fraction) of the
 reflector viewed from the plasma.
 \par\noindent
 \footnotemark[$\S$] Iron abundance in a unit of solar. Constrained to
 be common among the three energy bands. The solar
 abundance table of \citet{1989GeCoA..53..197A} is adopted, in which
 [Fe/H] = $4.68\times 10^{-5}$.
 \par\noindent
 \footnotemark[$\|$] Oxygen abundance in a unit of solar. Constrained to
 be common among the three energy bands. The solar
 abundance table of \citet{1989GeCoA..53..197A} is adopted, in which
 [O/H] = $8.51\times 10^{-4}$.
}\hss}}
\end{tabular}
\end{center}
\end{table}
The fits are marginally acceptable at the 90\% confidence level.
The parameters of the plasma and the reflector are determined consistently
between quiescence and outburst with unprecedented precision. We adopt
\citet{1989GeCoA..53..197A} as the solar abundances of the metals. The
DEM power $\alpha$ is varied independently between the iron and oxygen
bands in the outburst spectra, because the cooling functions are
significantly different between these two bands
\citep{1993ApJ...418L..25G}. In fact, the resultant $\alpha$ values are
significantly different between O ($\simeq -0.4$) and Fe
($\sim$6). Considering this large difference, we have checked mutual
consistency of the model spectra in the two energy bands in
outburst. The result is shown in Fig.~\ref{fig:consistency}.
The extrapolation of the best-fit model in one band does not exceed the
observed flux in the other band. Hence, the fits to the two energy bands
in outburst are revealed to be mutually consistent. The maximum
temperature of the plasma in quiescence ($T^{\rm Q}_{\rm max} =
20.4^{+4.0}_{-2.6}$~keV) is significantly higher than that in outburst
($T^{\rm O}_{\rm max} = 6.0^{+0.2}_{-1.3}$~keV). Our $T^{\rm Q}_{\rm
max}$ is consistent with that of \citet{1997MNRAS.288..649D} ($=
21^{+11}_{-5.7}$~keV).  Their $T^{\rm O}_{\rm max}$ ($=
9.6^{+3.4}_{-1.7}$~keV) seems slightly higher than ours, though this may
be due to different outburst phases at which the observations were
carried out.  The optically thin thermal plasma is covered with the
reflector more in quiescence ($\Omega^{\rm Q}/2\pi = 1.7\pm0.2$) than in
outburst ($\Omega^{\rm O}/2\pi =0.9^{+0.5}_{-0.4}$). The covering
fractions are not constrained by \citet{1997MNRAS.288..649D} very well
($\Omega^{\rm Q}/2\pi = 0.6^{+1.1}_{-0.6}$ and $\Omega^{\rm O}/2\pi =
2.2^{+1.8}_{-1.5}$). The iron and oxygen abundances are both sub solar
($Z_{\rm Fe} = 0.37^{+0.01}_{-0.03}Z_\odot$ and $Z_{\rm O} =
0.46^{+0.04}_{-0.03}Z_\odot$).

Finally, we have considered the effects of the 5\% systematic PIN NXB
error. We have made background spectra of the PIN with an intensity
being enhanced or reduced by 5\%, subtracted them from the quiescence
PIN spectrum, and repeated the combined spectral fit described
above. The resultant systematic errors are summarized in parentheses in
table~\ref{tab:simulfit}. We have found $T^{\rm Q}_{\rm max}$ is
accompanied by the largest fractional systematic error of
$\pm$3.0~keV. Fortunately, those of the other parameters are smaller
than the statistical ones. Since we do not use the outburst PIN data,
there appears no systematic error in $T^{\rm O}_{\rm max}$, $\alpha^{\rm
O}$, and $\Omega^{\rm O}/2\pi$.

\subsection{Abundance of the other elements \label{sec:abundances}}

\begin{figure}[htbp]
\begin{center}
\begin{minipage}{0.45\textwidth}
 \includegraphics[width=\textwidth,bb=0 0 979 746]{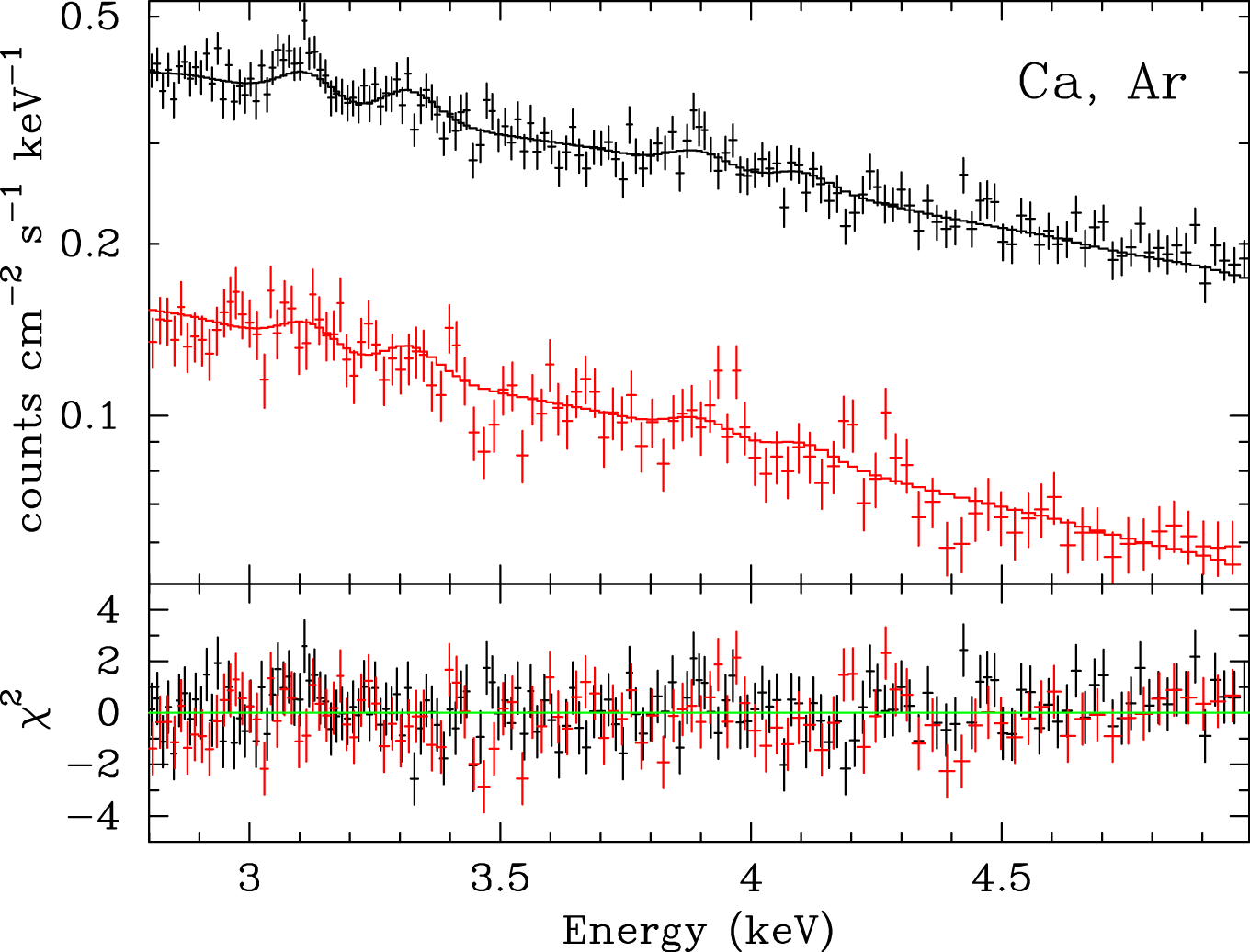}
\end{minipage}
\begin{minipage}{0.45\textwidth}
 \includegraphics[width=\textwidth,bb=0 0 979 746]{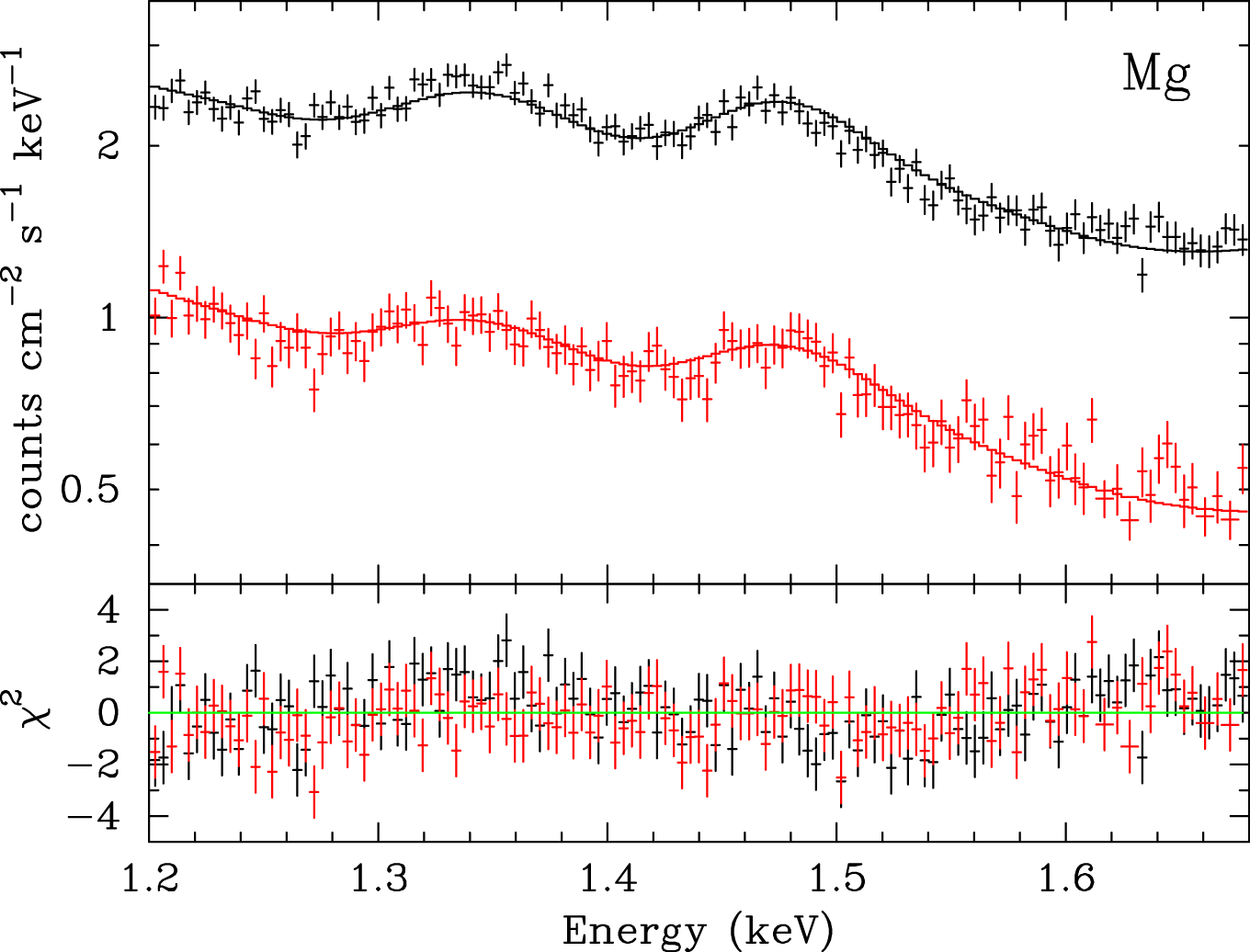}
\end{minipage}
\begin{minipage}{0.45\textwidth}
 \includegraphics[width=\textwidth,bb=0 0 979 746]{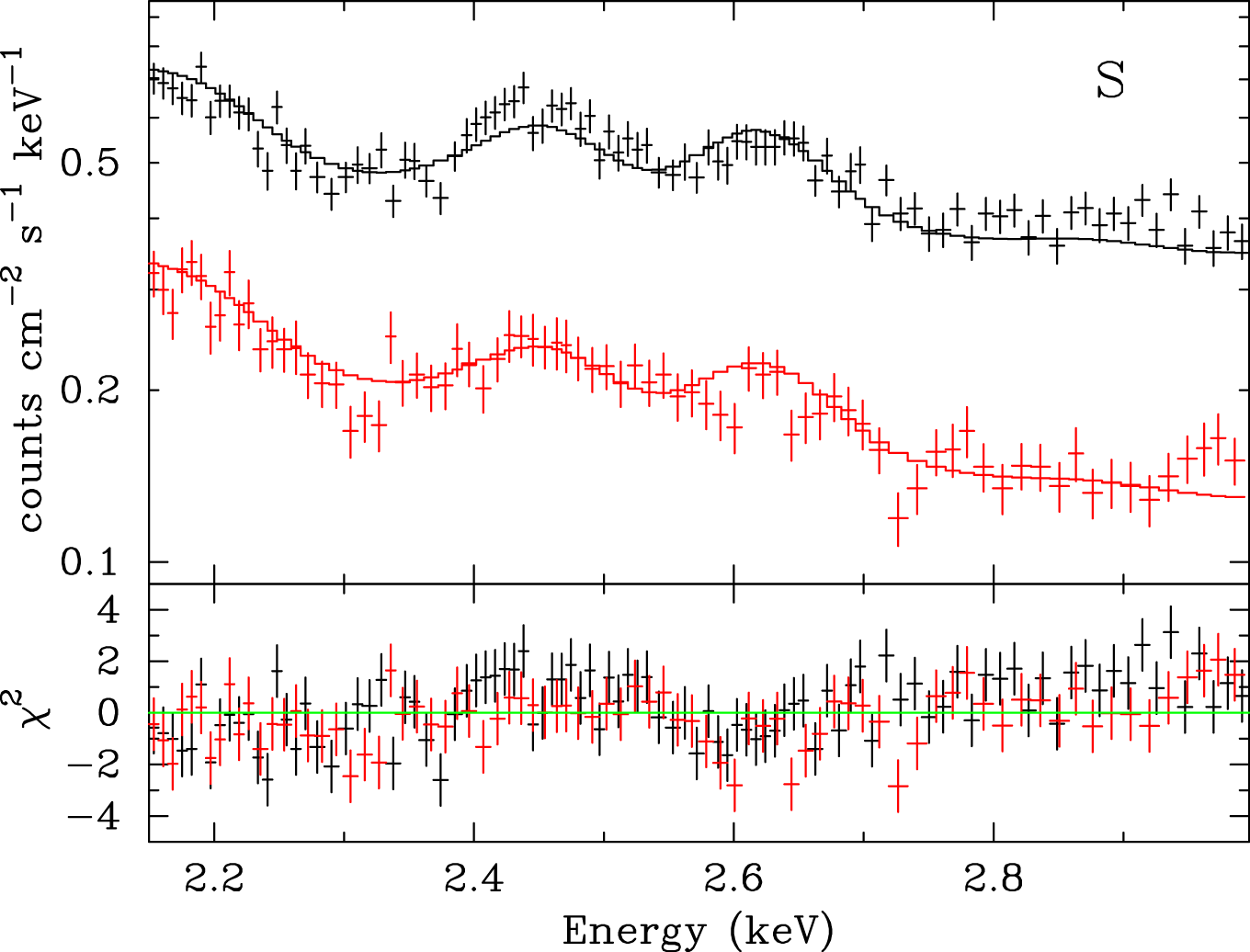}
\end{minipage}
\begin{minipage}{0.45\textwidth}
 \includegraphics[width=\textwidth,bb=0 0 979 746]{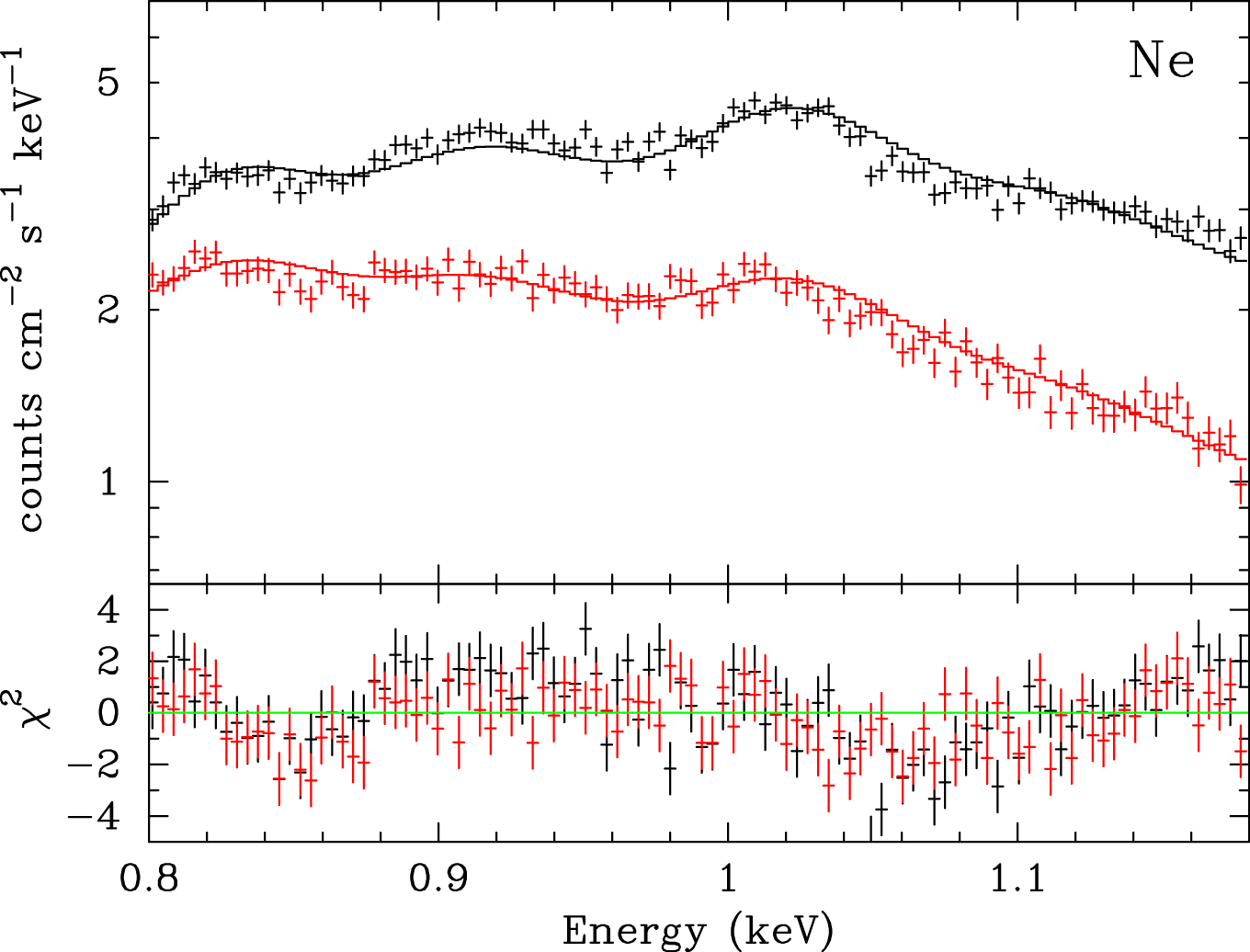}
\end{minipage}
\begin{minipage}{0.45\textwidth}
 \includegraphics[width=\textwidth,bb=0 0 980 747]{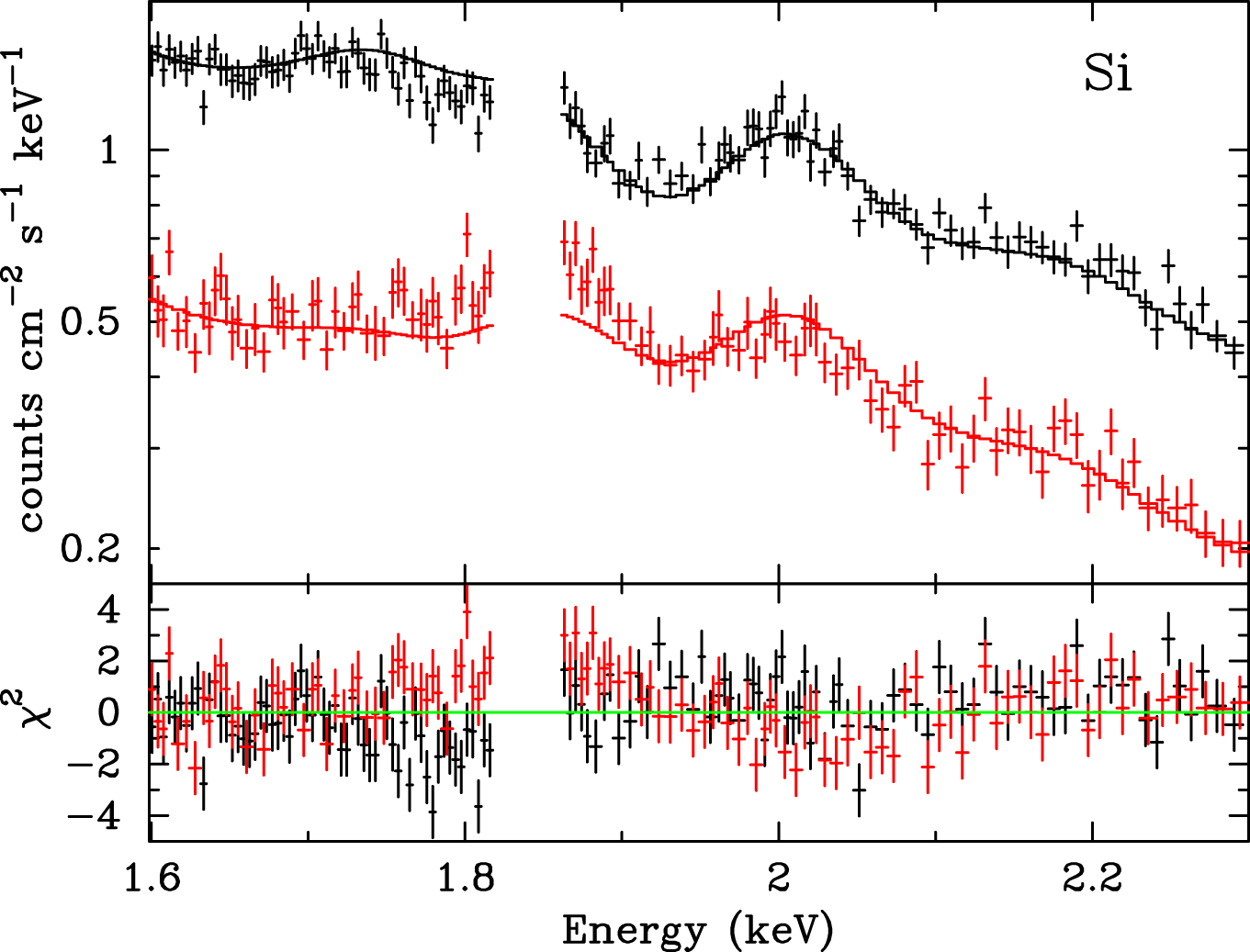}
\end{minipage}
\begin{minipage}{0.45\textwidth}
 \includegraphics[width=\textwidth,bb=0 0 979 757]{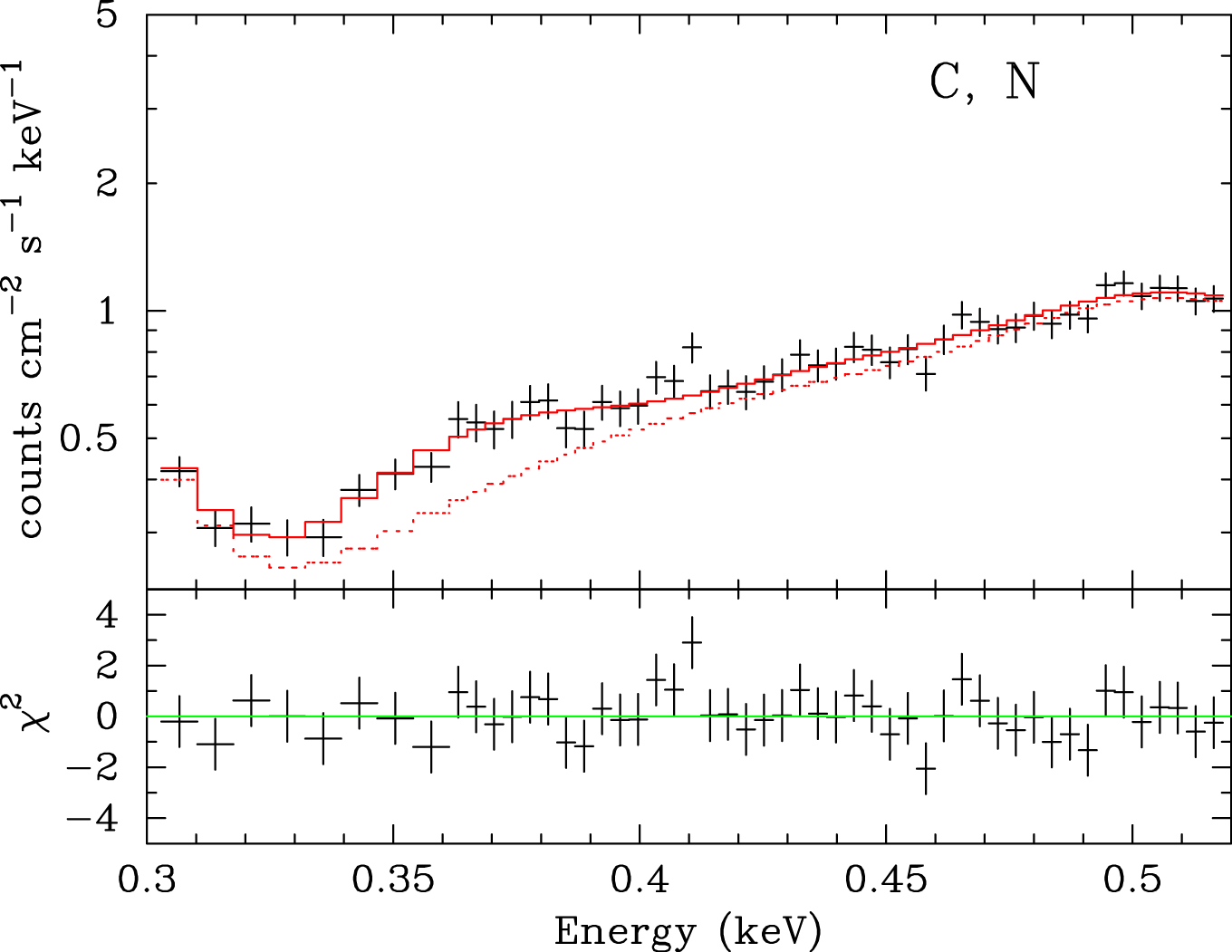}
\end{minipage}
\end{center}
\caption{Local {\tt cevmkl} fits to the He-like and H-like {\ka}
 emission lines of Ca, Ar, S, Si, Mg, Ne, N, and C. The upper panel
 shows the data (crosses) and the models (histograms), while the lower
 panel shows the residuals in a unit of the standard deviation. In order
 to show detection of the C emission lines, we have also plotted a {\tt
 cevmkl} model with the C abundance being set equal to zero as a dotted
 histogram in panel (f).  \label{fig:abundances}}
\end{figure}

We can identify He-like and H-like {\ka} emission lines from nitrogen to
iron in the outburst spectra shown in Fig.~\ref{fig:avespec}. Utilizing
the intensities of these lines, we can estimate the abundances of
corresponding elements. The line intensities depend upon $T^{\rm O}_{\rm
max}$, $\alpha^{\rm O}$, and the abundance. Among them, We have
constrained $T^{\rm O}_{\rm max}$ well by the combined fit analysis
described in \S~\ref{sec:plasmaparam}. With $T^{\rm O}_{\rm max}$ being
fixed at 6.0~keV (table~\ref{tab:simulfit}), we have evaluated the
abundances by fitting the {\tt cevmkl} model to the {\ka} lines of each
element separately. In principle, $\alpha^{\rm O}$ (see
eq.~(\ref{eq:dem})) is determined by the relative intensities of the
He-like and H-like {\ka} lines, and the abundance is constrained by and
their EWs. The results of the fits are shown in
Fig.~\ref{fig:abundances} and the best-fit abundances as well as
$\alpha^{\rm O}$ are summarized in table~\ref{tab:abundances}.
\begin{table}[htbp]
\setlength{\tabcolsep}{4pt}
\caption{Summary of the abundances of iron, calcium, argon, sulfur,
 silicon, magnesium, neon, oxygen, and carbon. }
\label{tab:abundances}
 \begin{center}
  \begin{tabular}{cccccccc} \hline \hline
Element & Energy Band  & $N_{\rm H}$\footnotemark[$\ddagger$]
 & $T^{\rm O}_{\rm max}$\footnotemark[$\dagger$] & $\alpha^{\rm O}$
 & Abundance & ${\rm \chi^2}$ (d.o.f.) & ${\rm \chi^2_{\nu}}$ \\
%%%%
 & (keV) & ($10^{19}$cm$^{-2}$) & (keV)
 & & (${\rm Z_\odot}$) 
 & & \\
\hline
Fe\footnotemark[\P]
 & 4.2--10 & ${\rm 5.0^{f}}$ & $6.0^{+0.2}_{-1.3}$
 & ${\rm 5.8^{+2.6}_{-1.4}}$ & ${\rm 0.37^{+0.01}_{-0.03}}$
 & 728.3 (657)
 & 1.11 \\
Ca
 & 2.8--5.0 & ${\rm 5.0^{f}}$ & ${\rm 6.0^{f}}$
 & ${\rm 2.2^{+0.5}_{-0.4}}$ & ${\rm 0.73^{+0.35}_{-0.36}}$
 & 263.7 (265) & 1.00 \\
Ar
 & & &
 & & ${\rm 0.93^{+0.34}_{-0.32}}$
 & & \\
S
 & 2.15--3.0 & ${\rm 5.0^{f}}$ & ${\rm 6.0^{f}}$
 & ${\rm 0.72^{+0.16}_{-0.14}}$ & ${\rm 0.88\pm{0.09}}$
 & 242.1 (171) & 1.42 \\
Si
 & 1.6--2.3 & ${\rm 5.0^{f}}$ & ${\rm 6.0^{f}}$
 & ${\rm 0.88^{+0.15}_{-0.13}}$ & ${\rm 0.90\pm{0.07}}$
 & 389.0 (244) & 1.59 \\
Mg & 1.2--1.7 & ${\rm 5.0^{f}}$ & ${\rm 6.0^{f}}$
 & ${\rm 0.44\pm0.06}$ & ${\rm 0.80\pm0.06}$
 & 316.2 (267) & 1.19 \\
Ne & 0.8--1.2
 & ${\rm 5.0^{f}}$ & ${\rm 6.0^{f}}$
 & ${\rm 0.31\pm0.02}$ & ${\rm 0.49\pm0.04}$
 & 429.8 (217) & 1.99 \\
\hline
\multicolumn{8}{c}{fit with the BI spectrum only}\\
\hline
O\footnotemark[\P]
 & 0.55--0.72 & ${\rm 5.0^{f}}$ & $6.0^{+0.2}_{-1.3}$
 & ${\rm -0.36^{+0.02}_{-0.03}}$ & ${\rm 0.46^{+0.04}_{-0.03}}$
 & 728.3 (657) & 1.11 \\
N
 & 0.3--0.52 & ${\rm 3.5^{f}}$  & ${\rm 6.0^{f}}$
 & ${\rm 0.68^{+0.20}_{-0.23}}$ & ${\rm 0.40^{+0.99}_{-0.40}}$
 & 44.7 (49) & 0.91 \\
C
 & & &
 & & ${\rm 3.47^{+1.60}_{-0.99}}$
 & & \\
N
 & 0.3--0.52 & ${\rm 5.0^{f}}$ & ${\rm 6.0^{f}}$
 & ${\rm 0.64^{+0.17}_{-0.22}}$ & ${\rm 0.30^{+0.97}_{-0.30}}$
 & 44.6 (49) & 0.91 \\
C
 & & &
 & & ${\rm 3.20^{+1.41}_{-0.90}}$
 & & \\
N & 0.3--0.52 & ${\rm 7.9^{f}}$ & ${\rm 6.0^{f}}$
 & ${\rm 0.56^{+0.10}_{-0.22}}$ & ${\rm 0.13^{+0.94}_{-0.13}}$
 & 44.6 (49) & 0.91 \\
C
 & & &
 & & ${\rm 2.76^{+1.14}_{-0.76}}$
 & & \\
\hline
\multicolumn{8}{@{}l@{}}{\hbox to 0pt{\parbox{140mm}{\footnotesize
Note.---\ `f' implies the parameter is fixed.
   \par\noindent
\footnotemark[$\dagger$] Fixed at 6.0~keV obtained from the simultaneous
   fit (table~\ref{tab:simulfit} in \S~\ref{sec:plasmaparam}). 
   \par\noindent
\footnotemark[$\ddagger$] Fixed at ${\rm 5.0\times10^{19}~cm^{-2}}$
   obtained from the Chandra LETG observation \citep{2004ApJ...610..422M}.
   \par\noindent
\footnotemark[\P] The same values as summarized in table~\ref{tab:simulfit}.
}\hss}}
\end{tabular}
\end{center}
\end{table}
The spectral fit is started from a higher energy band. This is because
the line emissions from lighter elements do not affect the determination
of the abundances of heavier elements, whereas L-shell emission from the
heavier elements is possible to affect the fit to the lighter element
lines. As the abundances of the heavier elements have been measured, those
of the lighter elements are determined one after the other by fixing the
abundances of the heavier elements at their best-fit values. The fit to
Ca and Ar is carried out simultaneously, and so is to N and C. Note that
only BI-CCD is used for the fit to N and C.

In order to estimate the abundances of N and C, we have to take into
account several possible systematic effects such as the low energy
response of the XIS, the uncertainty of the contamination accumulating
on the optical blocking filters above the CCD chips, and systematic
error of the hydrogen column density to SS~Cyg. Among them, the low
energy response and the contamination are investigated by ourselves with
the PKS~2155--304 data taken during Nov. 30 -- Dec. 2, 2005, which date
is close enough to our SS~Cyg observations. As summarized in
Appendix~\ref{appendix:A}, we need to apply additional carbon K-edge
with $\tau = 0.88\pm 0.05$ and extra $N_{\rm H} = 8.2\pm 0.7\times
10^{19}$~cm$^{-2}$. By applying these corrections to the model, we can
utilize the BI-CCD data down to 0.23~keV. In order to estimate the
$N_{\rm H}$ systematic error, on the other hand, we have checked
variation of the best-fit parameters with an $N_{\rm H}$ of 3.5, 5.0,
and 7.9$\times 10^{19}$~cm$^{-2}$. These values are selected based on
$N_{\rm H} = 5.0^{+2.9}_{-1.5}\times 10^{19}$~cm$^{-2}$ obtained from
the Chandra LETG observation of SS~Cyg in outburst
\citep{2004ApJ...610..422M}.  Since the H-like {\ka} line of C and the
He-like {\ka} line of N cannot be resolved with the BI-CCD energy
resolution, we employ the BI-CCD spectrum in the 0.3--0.52~keV band
covering the emission lines of both C and N, and fit {\tt cevmkl} model.

We remark that not all the fits shown in Fig.~\ref{fig:abundances} are
acceptable. The fit to Si is especially poor. As evident from the fit
residuals inconsistent between the BI and FI-CCDs at around the Si
K-edge (1.7--1.9~keV), this is mainly due to systematic error associated
with energy response calibration. Nevertheless, the best-fit parameters
summarized in table~\ref{tab:abundances} indicate that the abundances of
the medium-Z elements (Si, S, and Ar) are close to solar ones, whereas
they are reduced for both lighter and heavier elements. The exception is
the carbon abundance whose lower limit is 2.0$Z_\odot$ even if we
consider the systematic error of $N_{\rm H}$. Since the
energy resolution of the XIS in the C line energy band is not so good as
in higher energy band, we have drawn the {\tt cevmkl} model with the C
abundance being set equal to zero in Fig.~\ref{fig:abundances}(f) as the
dotted histogram. The data excess associated with the C emission is
clear.

Given the spectral parameters, we can calculate X-ray fluxes of SS~Cyg,
which is $19.5\times10^{-11}$~erg~cm$^{-2}$~s$^{-1}$ and $8.1\times
10^{-11}$~erg~cm$^{-2}$~s$^{-1}$ in the 0.4--10~keV band in quiescence
and outburst, respectively. Assuming the distance of 166~pc
\citep{1999ApJ...515L..93H}, we obtain the luminosities of $6.4\times
10^{32}$~erg~s$^{-1}$ and $2.7\times 10^{32}$~erg~s$^{-1}$,
respectively, in this energy band.

\subsection{Iron emission lines}

\subsubsection{Quiescence \label{sec:FeinQ}}

\begin{figure}[htb]
 \includegraphics[width=0.49\textwidth,bb=0 0 976 750]{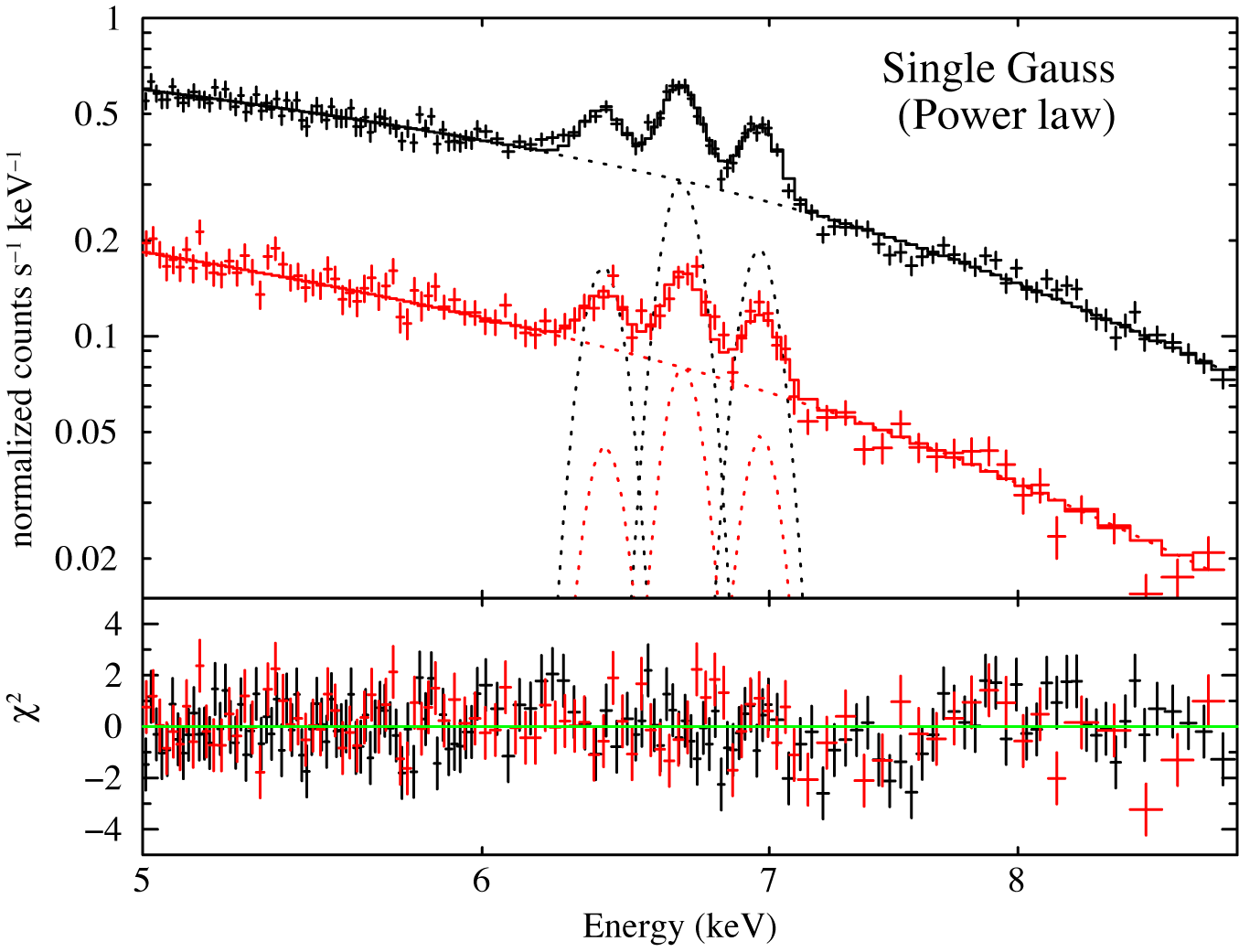}
 \includegraphics[width=0.49\textwidth,bb=0 0 975 750]{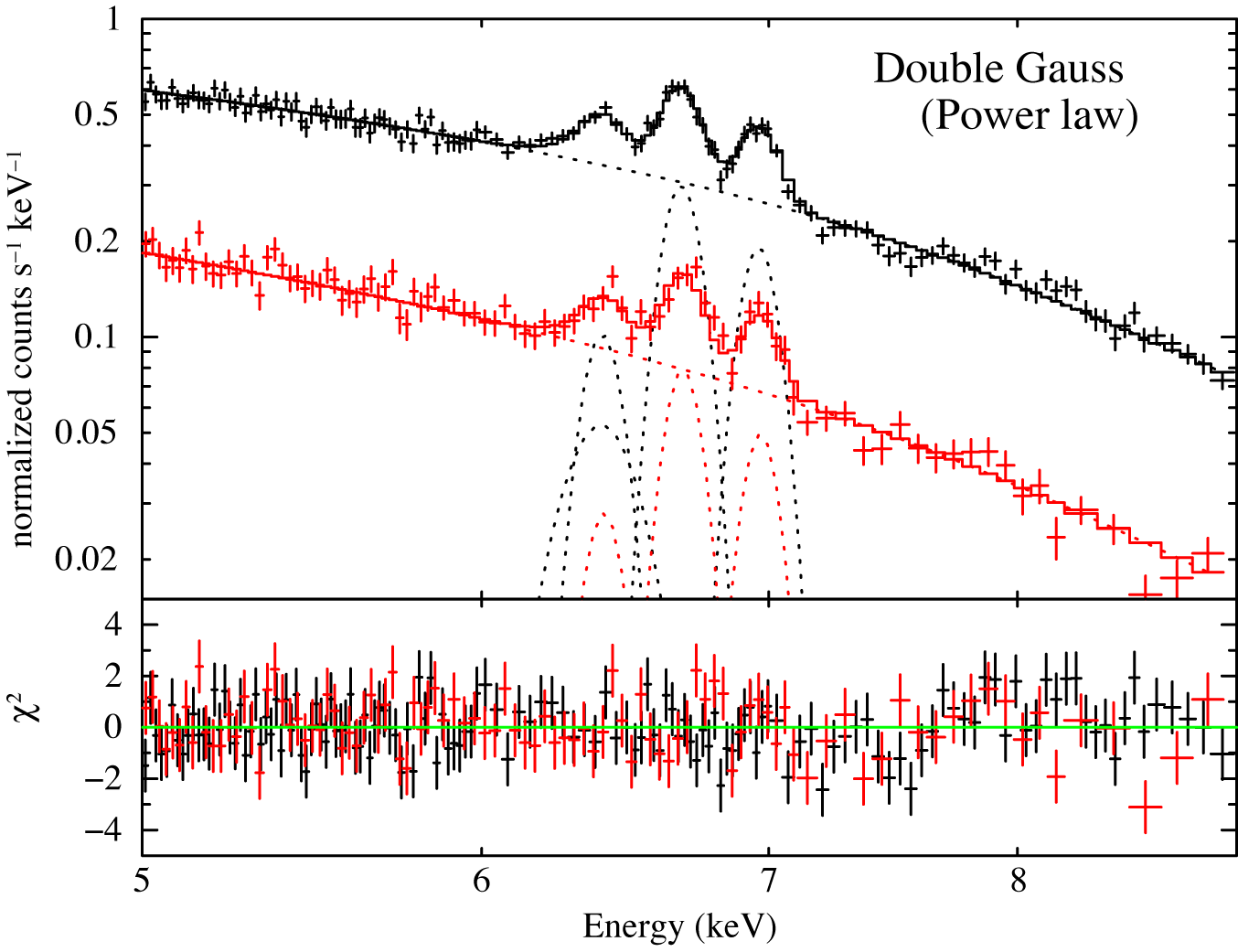}
 \caption{Fits of a power law plus 3 Gaussians (left) and 4 Gaussians
 (right) to the 5--9~keV XIS spectra in quiescence. \label{fig:linefit}}
\end{figure}

In order to constrain the geometry of the optically thin thermal plasma
in quiescence, we have attempted to evaluate the parameters of the
6.4~keV line in detail. We fit a model composed of a power-law continuum
and three narrow Gaussian lines at 6.4~keV, 6.7~keV and 7.0~keV to the
XIS spectra in the 5--9~keV band. The result of the fit is shown in
Fig.~\ref{fig:linefit} and the best-fit parameters are summarized
in table~\ref{tab:linefit}.
\begin{table}[htb]
\caption{Best-fit parameters of the iron {\ka} emission lines with the
 5--9~keV XIS spectra in quiescence. \label{tab:linefit}}
 \begin{center}
  \begin{tabular}{llll}
\hline\hline
Component & Parameter & 3 lines & 4 lines \\ \hline
Continuum & $\Gamma$\footnotemark[$\|$]
                           & $1.60^{+0.03}_{-0.02}$
                           & $1.62^{+0.02}_{-0.03}$\\
 & $N_{\rm C}$\footnotemark[$\dagger$] & $9.13\pm0.07$ & $9.50\pm0.08$ \\
H-like line & $E_{\rm L}$\footnotemark[$\ast$] (keV)
                           & $6.968^{+0.003}_{-0.008}$
                           & $6.966^{+0.006}_{-0.008}$ \\
 & $N_{\rm L}$\footnotemark[$\ddagger$] 
 & $5.30^{+0.39}_{-0.41}$  & $5.39^{+0.38}_{-0.42}$ \\
 & $\sigma$ (keV) & 0 (fixed) & 0 (fixed) \\
 & EW\footnotemark[$\S$] (eV) & $130^{+9}_{-10}$ & $132^{+9}_{-10}$ \\
He-like line & $E_{\rm L}$\footnotemark[$\ast$] (keV)
                           & $6.682^{+0.002}_{-0.008}$ 
                           & $6.684^{+0.004}_{-0.006}$ \\
 & $N_{\rm L}$\footnotemark[$\ddagger$] 
 & $7.81^{+0.41}_{-0.42}$ & $7.67^{+0.36}_{-0.48}$ \\
 & $\sigma$ (keV) & 0 (fixed) & 0 (fixed) \\
 & EW\footnotemark[$\S$] (eV) & $181\pm10$ & $175^{+8}_{-11}$ \\ \hline
\multicolumn{4}{c}{Fluorescence line parameters} \\ \hline
Narrow line & $E_{\rm L}$\footnotemark[$\ast$] (keV)
       & $6.407^{+0.008}_{-0.011}$ & $6.404^{+0.012}_{-0.008}$ \\
       & $N_{\rm L}$\footnotemark[$\ddagger$] 
       & $4.02^{+0.37}_{-0.36}$  & $2.46^{+0.91}_{-0.38}$ \\
       & $\sigma$ (keV) & 0 (fixed) & 0 (fixed) \\
       & EW\footnotemark[$\S$] (eV) & $86\pm8$ & $53^{+20}_{-9}$ \\
Broad line  & $N_{\rm L}$\footnotemark[$\ddagger$] 
       & ---  & $2.38^{+0.98}_{-0.50}$ \\
       & $\sigma$ (keV) & --- & $0.112^{+0.055}_{-0.031}$ \\
       & EW\footnotemark[$\S$] (eV) & ---  & $51^{+21}_{-11}$ \\ \hline
 & $\chi^2$ (d.o.f) & 261.7 (233) & 251.5 (231) \\
 & $\chi^2_{\nu}$ & 1.12 & 1.09 \\
\hline
\multicolumn{4}{@{}l@{}}{\hbox to 0pt{\parbox{85mm}{\footnotesize
\footnotemark[$\ast$] Line central energy. The value is constrained to
   be the same between the narrow and broad 6.4~keV components.
\par\noindent
\footnotemark[$\dagger$] Continuum normalization in a unit of
   $10^{-3}$~cm$^{-2}$~keV$^{-1}$~s$^{-1}$ at 1~keV.
\par\noindent
\footnotemark[$\ddagger$] Line normalization (intensity) in a unit of
   $10^{-5}$~photons~cm$^{-2}$~s$^{-1}$. 
\par\noindent
\footnotemark[$\S$] Equivalent width.
\par\noindent
\footnotemark[$\|$] Photon index.
  }\hss}}
\end{tabular}
\end{center}
\end{table}
There remains a slight excess emission in the low energy tail of the
6.4~keV line. We thus have added a broad Gaussian to the 6.4~keV line.
The results are also summarized in table~\ref{tab:linefit} and
Fig.~\ref{fig:linefit}. The fit is improved with a $\chi^2$ (d.o.f)
value from 261.7 (233) to 251.5 (231). The chance probability for this
to occur is 0.009, which implies adding a broad Gaussian is significant
at $\sim$99\% confidence level. The resultant equivalent width of the
narrow and broad lines is ${\rm EW\,(narrow)} = 53^{+20}_{-9}$~eV and
${\rm EW\,(broad)} = 51^{+21}_{-11}$~eV.

\subsubsection{Outburst \label{sec:FeinO}}

\begin{figure}[htb]
\centerline{
 \includegraphics[width=0.60\textwidth,bb=0 0 1038 734]{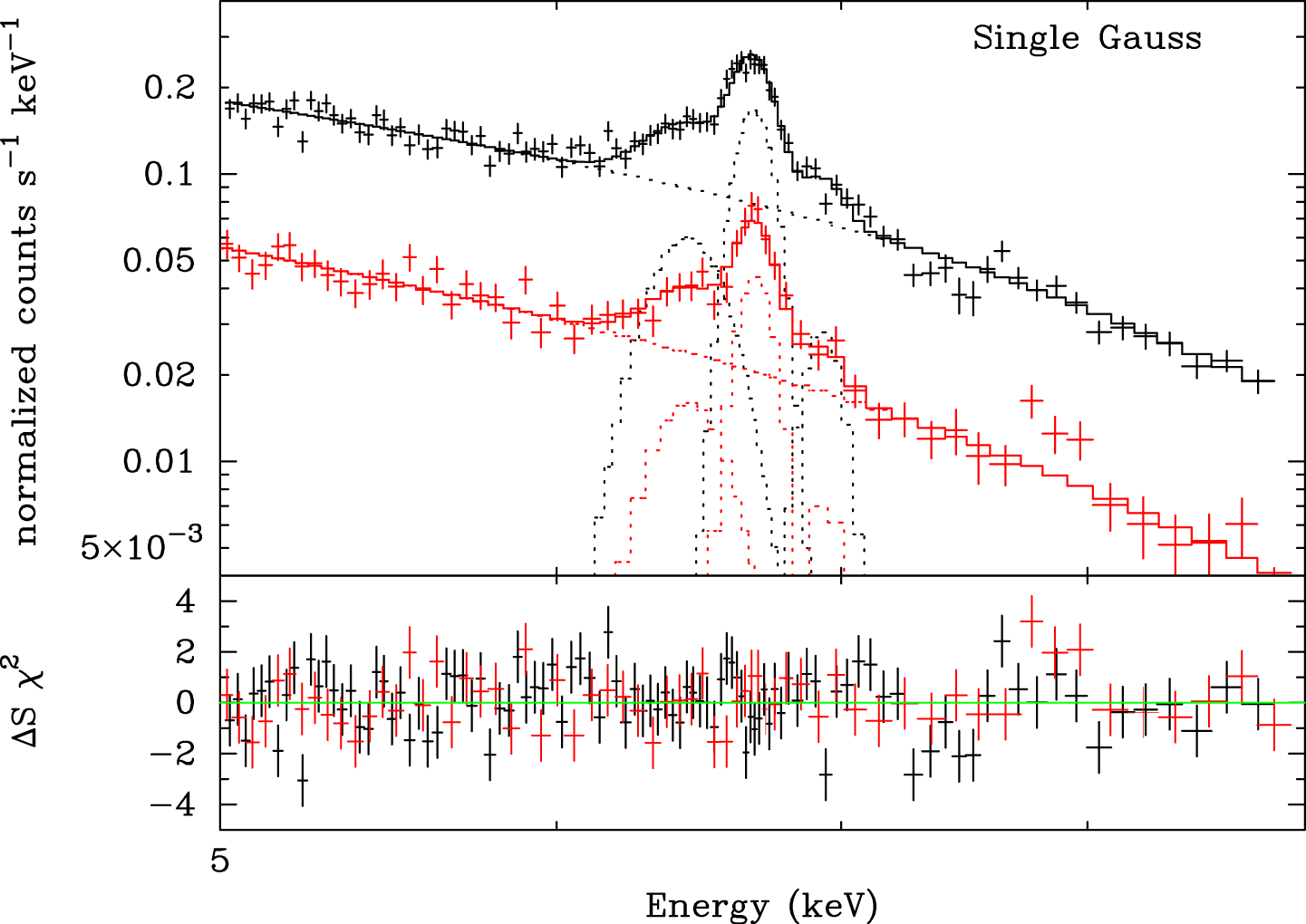}
}
 \caption{Fits of a power law plus 3 Gaussians to the 5--9~keV XIS spectra in outburst. \label{fig:linefit_o}}
\end{figure}

In outburst, the 6.4~keV line is obviously broad
(Fig.~\ref{fig:avespec}). We therefore adopt a broad Gaussian for
modeling the 6.4~keV emission line. As in the case of the fit to the
quiescence spectra (\S~\ref{sec:FeinQ}), we have tried to fit a model
composed of a power law and three Gaussians to the XIS spectra in the
5--9~keV band. The result is shown in Fig.~\ref{fig:linefit_o}, and the
best-fit parameters are summarized in table~\ref{tab:linefit_o}.
\begin{table}[htb]
\caption{Best-fit parameters of the iron {\ka} emission lines with the
 5--9~keV XIS spectra in outburst. \label{tab:linefit_o}}
 \begin{center}
  \begin{tabular}{lll}
\hline\hline
Component & Parameter & \\ \hline
Continuum & $\Gamma$\footnotemark[$\|$]              
       &   $2.16^{+0.09}_{-0.05}$  \\
 & $N_{\rm C}$\footnotemark[$\dagger$]               
       &   $6.77^{+0.65}_{-0.12}$  \\
H-like {\ka} & $E_{\rm L}$\footnotemark[$\ast$] (keV) 
       & $6.936^{+0.028}_{-0.036}$ \\
 & $N_{\rm L}$\footnotemark[$\ddagger$]   
       &  $0.82\pm 0.17$   \\
 & $\sigma$ (keV)                         
       & 0 (fixed)                 \\
 & EW\footnotemark[$\S$] (eV)             
       & $80\pm 17$ 	 \\
He-like {\ka} & $E_{\rm L}$\footnotemark[$\ast$] (keV)
       & $6.682^{+0.004}_{-0.007}$ \\
 & $N_{\rm L}$\footnotemark[$\ddagger$]   
       &   $4.37^{+0.29}_{-0.30}$  \\
 & $\sigma$ (keV)                         
       & 0 (fixed)	         \\
 & EW\footnotemark[$\S$] (eV)             
       & $390^{+26}_{-27}$ \\
Neutral {\ka}& $E_{\rm L}$\footnotemark[$\ast$] (keV) 
       & $6.448^{+0.024}_{-0.025}$\\
 & $N_{\rm L}$\footnotemark[$\ddagger$]   
       & $2.90^{+0.95}_{-0.38}$    \\
 & $\sigma$ (keV)                         
       & $0.12^{+0.06}_{-0.03}$    \\
 & EW\footnotemark[$\S$] (eV)             
       & $208^{+68}_{-27}$ \\ \hline
 & $\chi^2$ (d.o.f) & 198.1 (155) \\
 & $\chi^2_{\nu}$ & 1.28 \\
\hline
\multicolumn{3}{@{}l@{}}{\hbox to 0pt{\parbox{85mm}{\footnotesize
\footnotemark[$\ast$] Line central energy. The value is constrained to
   be the same between the narrow and broad 6.4~keV components.
\par\noindent
\footnotemark[$\dagger$] Continuum normalization in a unit of
   $10^{-3}$~cm$^{-2}$~keV$^{-1}$~s$^{-1}$ at 1~keV.
\par\noindent
\footnotemark[$\ddagger$] Line normalization (intensity) in a unit of
   $10^{-5}$~photons~cm$^{-2}$~s$^{-1}$. 
\par\noindent
\footnotemark[$\S$] Equivalent width.
\par\noindent
\footnotemark[$\|$] Photon index.
  }\hss}}
\end{tabular}
\end{center}
\end{table}
The equivalent width of the 6.4~keV line is as large as $\sim$210~eV.
The energy width of the line $\sigma = 0.12$~keV is nearly the same as
that of the broad component in quiescence.

In quiescence, the narrow component dominates the 6.4~keV line. In order
to see if there is a narrow component also in the outburst spectra, we
have implemented another narrow Gaussian ($\sigma =0$) into the model at
the same central energy of the broad 6.4~keV component. The fit with
this model, however, only provides an upper limit to the narrow 6.4~keV
line with an equivalent width of $<39$~eV, or $<20$\% of the broad
component.

%%%%%%%%%%%%%%%%%%%%
\section{Discussion}
%%%%%%%%%%%%%%%%%%%%

\subsection{Location of the optically thin thermal plasma in quiescence}

\subsubsection{Geometry of the plasma \label{sec:BLGeoinQ}}

\begin{figure}[htb]
\centerline{
 \includegraphics[width=0.6\textwidth,bb=0 0 1068 745]{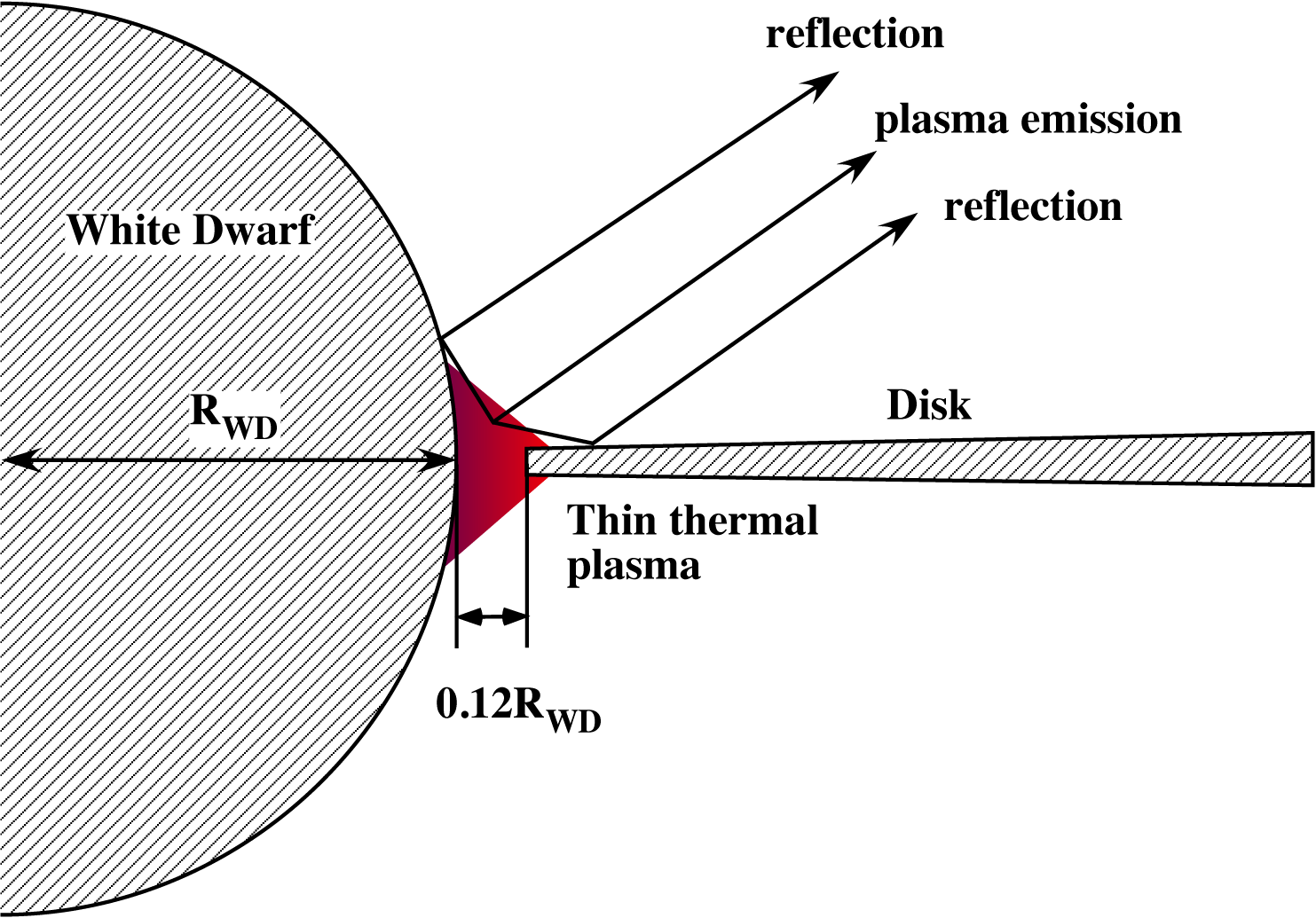}
}
\caption{Geometry of the BL in quiescence deduced from the
 Suzaku observation. \label{fig:BLGeoinQ}}
\end{figure}

We have shown in \S~\ref{sec:plasmaparam} that the optically thin
thermal plasma in quiescence is covered with the reflector with a solid
angle of $\Omega^{\rm Q}/2\pi=1.7\pm0.2$. Such a high value of
$\Omega^{\rm Q}/2\pi$ is achieved only in a limited number of high-mass
X-ray binaries (e.g. \cite{2003ApJ...597L..37W}) that are surrounded by
matter as thick as $\sim$10$^{24}$~H~cm$^{-2}$. The thickness of matter
to SS~Cyg, on the other hand, only amounts to $N_{\rm H} \lesssim
10^{20}$~cm$^{-2}$ or $\tau \lesssim 10^{-4}$ for Thomson scattering,
which is too transparent to reflect X-rays efficiently. Candidates of
the reflection site are therefore limited to the white dwarf and the
accretion disk. The observed $\Omega^{\rm Q}/2\pi$, however, requires a
special geometry, because even an infinite slab can only subtend a solid
angle of $\Omega/2\pi = 1$ over an X-ray source above it. The plasma
should be compact enough compared to the size of the white dwarf, and
should be located very close to the white dwarf and the accretion
disk. We consider the standard optically thin BL
\citep{1985ApJ...292..535P} as the most plausible configuration of the
optically thin thermal plasma in quiescence (see also
Fig.~\ref{fig:BLGeoinQ}).
In \S~\ref{sec:FeinQ}, we have shown that the 6.4~keV iron line profile
favors the broad component in addition to the narrow component, which
can naturally be interpreted as originating from the accretion disk and
the white dwarf, respectively, via fluorescence. The width of the broad
component $\sigma = 0.11^{+0.06}_{-0.03}$~keV corresponds to a
line-of-sight velocity dispersion of 5300$^{+2500}_{-1500}$~km~s$^{-1}$,
which is consistent with the line-of-sight Keplerian velocity amplitude
of the accretion disk just on the 1.19$M_\odot$ white dwarf ($v_{\rm
K}\sin i =$ 3800$\pm$400~km~s$^{-1}$, where we set $i=37^\circ \pm
5^\circ$).

\subsubsection{Size of the boundary layer \label{sec:sizeBL} }

Now that we know the abundance of iron $0.37^{+0.01}_{-0.03}Z_\odot$
(\S~\ref{sec:plasmaparam}), we can estimate the height of the boundary
layer in quiescence on the basis of the standard BL geometry displayed
in Fig.~\ref{fig:BLGeoinQ}. \citet{1991MNRAS.249..352G} theoretically
calculated the equivalent width of the fluorescent iron {\ka} line due
to a point source being located above an infinite slab (i.e,
$\Omega\!=\!2\pi$). The equivalent width depends on the inclination
angle $i$ between the observer's line of sight and the normal of the
slab, a photon index of the spectrum of the illuminating point source,
and an iron abundance of the slab. In applying their calculation to the
observed narrow 6.4~keV component, we have adopted the following
parameters;
\begin{enumerate}
\item 
We set the inclination angle $i\!=\!60^{\circ}$, which is the average
over the visible hemisphere of the white dwarf surface
\citep{1997MNRAS.288..649D}.
\item 
The observed continuum spectra above 5~keV are represented with a power
law with a photon index of $\sim$1.6
(table~\ref{tab:linefit}). This photon index is, however, affected by
the reflected continuum. In order to correct this, we have retried a fit
with a power law plus its reflection with $\Omega^{\rm Q}/2\pi$ and
$Z_{\rm Fe}$ fixed at the values listed in table~\ref{tab:simulfit}.
The resultant photon index is $\Gamma = 1.8$, which we adopt for the
power law incident on the white dwarf surface.
\end{enumerate}
With this parameter set, we have obtained an expected equivalent width
of the narrow 6.4~keV iron line to be 110~eV, according to Fig.~14 of
\citet{1991MNRAS.249..352G}. Note here that this figure is drawn under
the condition of the solar abundance with a composition of [Fe/H] =
$3.2\times 10^{-5}$ whereas our case is 0.37$Z_\odot$ under the
condition of [Fe/H] = $4.68\times 10^{-5}$
\citep{1989GeCoA..53..197A}. After correcting this abundance difference
using their Fig.~16, the expected equivalent width of the narrow 6.4~keV
line is
\begin{equation}
{\rm EW_{expected}} \;=\; 80\,\left( \frac{\Omega_{\rm WD}}{2\pi}\right)
 \;\;\;\mbox{[eV]},
\end{equation}
where $\Omega_{\rm WD}$ is the solid angle of the white dwarf viewed
from the BL plasma. Equating this to the observed ${\rm EW} =
53^{+20}_{-9}$ (table~\ref{tab:linefit}), we obtain $\Omega_{\rm
WD}/2\pi = 0.66^{+0.25}_{-0.11}$.  By assuming that the plasma is
point-like and is located at a height $h$ above the white dwarf, we can
directly link $\Omega_{\rm WD}/2\pi$ with $h$. From a simple geometrical
consideration, we obtain $h\!=\!0.060^{+0.055}_{-0.056}$ or $h <
0.12R_{\rm WD}$. Note that this number is similar to the thickness of
the BL estimated in the eclipsing dwarf nova HT~Cas $h < 0.15R_{\rm WD}$
\citep{1997ApJ...475..812M}. The total solid angle $\Omega^{\rm Q}/2\pi$
can be evaluated also from the 6.4~keV line by comparing the total
equivalent width of EW = $104^{+41}_{-20}$ to eq.~(3) as $\Omega/2\pi =
1.3^{+0.5}_{-0.2}$. This is consistent with $\Omega^{\rm Q}/2\pi =
1.7\pm0.2$ estimated by the continuum spectra. We therefore conclude
that the Suzaku results in quiescence favor the geometry of the standard
optically thin BL \citep{1985ApJ...292..535P} and its scale height from
the white dwarf is $h < 0.12R_{\rm WD}$.

\subsection{Location of the thin thermal plasma in outburst}
\subsubsection{Geometry of the plasma \label{sec:BLGeoinO}}
\begin{figure}[htb]
\centerline{
\includegraphics[width=0.7\textwidth,bb=0 0 747 258]{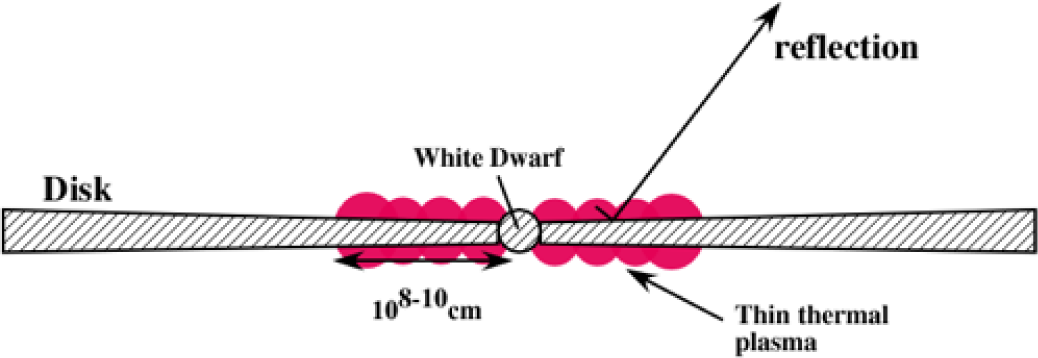}
}
\caption{Geometry of the optically thin thermal plasma in
outburst. The energy width of the 6.4~keV line, the small contribution
from the white dwarf surface to the EW, and $\Omega^{\rm O}/2\pi =
0.9^{+0.5}_{-0.4}$ strongly suggest that the plasma is located on the
disk like coronae. \label{fig:GeoinO}}
\end{figure}

As presented in \S~\ref{sec:FeinO}, the 6.4~keV iron line in outburst is
broad with a Gaussian $\sigma$ of 0.12~keV (table~\ref{tab:linefit_o}).
Since this width can be interpreted as the line-of-sight Kepler velocity
amplitude of the accretion disk on the white dwarf
(\S~\ref{sec:BLGeoinQ}), the main reflector in outburst is likely to be
the accretion disk. From the upper limit of the narrow 6.4~keV line,
contribution from the static white dwarf surface evident in quiescence
is negligible with an upper limit of 20\% of the disk contribution.
Moreover, the solid angle of the reflector $\Omega^{\rm O}/2\pi =
0.9^{+0.5}_{-0.4}$ is consistent with an infinite plane, reminiscent of
the disk, although the error is large. All these facts strongly suggest
that the plasma is located above the disk like coronae with their height
small enough compared with the disk radius, as shown in
Fig.~\ref{fig:GeoinO}.
This picture is supported also by the Chandra HETG observations
\citep{2008ApJ...680..695O}, in which the plasma emission lines are very
broad with their Gaussian $\sigma$ as large as
$\sim$2000~km~s$^{-1}$. This is much broader than the thermal velocity,
and can be attributed to the plasma being anchored to the disk by, for
example, magnetic field and corotating with the disk. A similar disk
corona geometry is suggested to explain the X-ray spectrum and the
energy width of the emission lines in WZ~Sge in outburst
\citep{2005ASPC..330..257W}.

Note that the absence of the narrow 6.4~keV component can alternatively
be understood by invoking the accretion belt
\citep{1978necb.conf...89P,1978A&A....63..265K} which is an equatorial
part of the white dwarf atmosphere being accelerated by the accretion
torque and rotating nearly at the local Keplerian velocity ($v_{\rm
K}\sin i\sim$4000~km~s$^{-1}$ in SS~Cyg, \S~\ref{sec:BLGeoinQ}).  It is
suggested from observations that the belt covers significant fraction of
the white dwarf surface in outburst
\citep{1993ApJ...405..327L,1997AJ....114.1165C,1998ApJ...497..928S}. If
so, the 6.4~keV line from the white dwarf should also be broad. Our data
of the 6.4~keV line in outburst can be compatible with the existence of
the accretion belt.

\subsubsection{A possible solution to the large EW of the 6.4 keV line}

As explained in \S~\ref{sec:BLGeoinO}, the width of the 6.4~keV line can
be attributed to the Keplerian motion of the disk (and the accretion
belt). The observed equivalent width $EW\!=\!208^{+68}_{-27}~{\rm eV}$
(table~\ref{tab:linefit_o}), however, can not be explained solely by
simple fluorescence due to continuum illumination. Assuming the
inclination angle of the disk $i=37^\circ$ \citep{1983PhDT........14S},
the photon index of the incident spectrum $\Gamma = 2.3$ (correction is
made to the value 2.16 in table~\ref{tab:linefit_o}, see
\S~\ref{sec:sizeBL}), the iron abundance of 0.37$Z_\odot$, we expect
${\rm EW} \sim 110$~eV at most according to \citet{1991MNRAS.249..352G},
even if we assume the upper limit of the allowed range of the solid
angle $\Omega^{\rm O}/2\pi = 1.4$, which may be achieved by taking the
accretion belt into account. Since the thickness of matter surrounding
SS~Cyg is thin, only of order $N_{\rm H} \lesssim 10^{20}$~cm$^{-2}$
(\S~\ref{sec:BLGeoinQ}), there is no other source that can produce
6.4~keV line photons via fluorescence. The large observed equivalent
width therefore requires some other mechanism to enhance the intensity
around 6.4~keV.

One possible solution is to invoke a Compton shoulder which appears in a
lower energy side of a main line due to Compton down scattering. Such a
shoulder is resolved with the Chandra HETG in a wide variety of sources
such as a HMXB \citep{2003ApJ...597L..37W}, a ULX
\citep{2002A&A...396..793B}, and an AGN \citep{2002ApJ...574..643K}.
Note that the seed photon of the shoulder in these objects is the
6.4~keV fluorescence line born in Compton-thick clouds, whereas in the
current case, we consider the He-like {\ka} line at 6.67~keV, because it
is very strong with an equivalent width of 390~eV
(table~\ref{tab:linefit_o}). The Compton shoulder of the 6.67~keV line
extends down to $\lambda_0 + 2\lambda_{\rm C}$ in wavelength,
corresponding to 6.50~keV in energy, in the case of the back scattering,
where $\lambda_0$ and $\lambda_{\rm C}$ are the wavelength of the
6.67~keV photon and the Compton wavelength (= $h/m_{\rm e}c$),
respectively. The low energy end of the shoulder (6.50~keV) is close
enough to be merged into the broad 6.4~keV line with a CCD resolution.

\begin{figure}[htb]
\centerline{
 \includegraphics[width=0.6\textwidth,bb=0 0 1020 750]{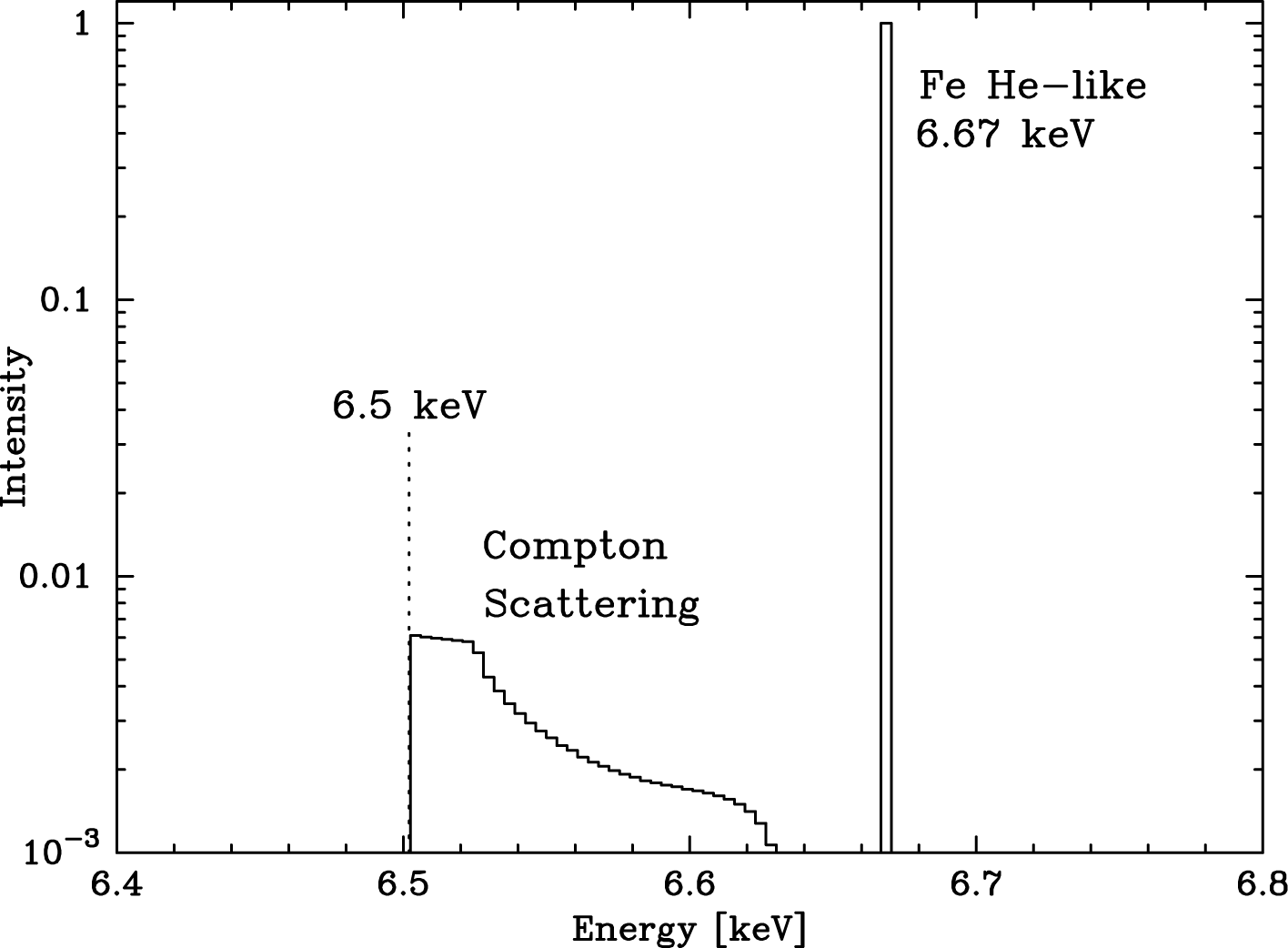}
 } 
\caption{Example of the line profile after a single Compton scattering
 by electrons in the geometrically thin non-relativistic disk. The
 energy of the source photons is 6.67~keV, which we consider a He-like
 iron {\ka} line. The inclination angle of the disk is
 37$^{\circ}$. The height of the photon source is $h=0.1R_{\rm out}$
 above the disk plane. \label{fig:comp_model_spec}}
\end{figure}

In order to evaluate contribution of the 6.67~keV Compton shoulder to
the observed EW of the 6.4~keV line, we consider a simple geometrical
model, in which a point source of 6.67~keV photons is centered at a
height $h$ above a circular slab with a radius of $R_{\rm out}$. We
adopt an inclination of the slab to be $i=37^\circ$
\citep{1983PhDT........14S}. Then, the energy of a scattered photon
arriving at the observer from any part of the slab can be calculated,
because the energy of a scattered photon is uniquely determined by the
scattering angle. In calculating the photon energy, we can ignore the
Kepler motion of the slab because the situation is in a non-relativistic
domain. Assuming that the intrinsic line emission is isotropic, we have
calculated the spectrum of the Compton-scattered photons by integrating
the emissions from the entire slab. Note that the intensity of the
spectrum is affected by the elemental abundance of the slab through
probability of the scattering (vs. photoelectric absorption), and
absorption of the scattered photons along the path to escape from the
slab. We adopt the abundances which we have already measured
(\S~\ref{sec:abundances}). As a result, the cross section of the
photoelectric absorption at 6.67~keV is obtained to be
$1.15\times10^{-24}\ {\rm cm^{-2}}$, which is roughly twice as large as
that of the Compton scattering. The resultant iron 6.67~keV line
spectrum in the case of $h = 0.1R_{\rm out}$ is shown in
Fig.~\ref{fig:comp_model_spec}.
The low energy end of the shoulder appears at 6.50~keV as expected, and
the high energy end is determined by the edge of the accretion disk.

\begin{figure}[htb]
\centerline{
 \includegraphics[width=0.6\textwidth,bb=0 0 1024 729]{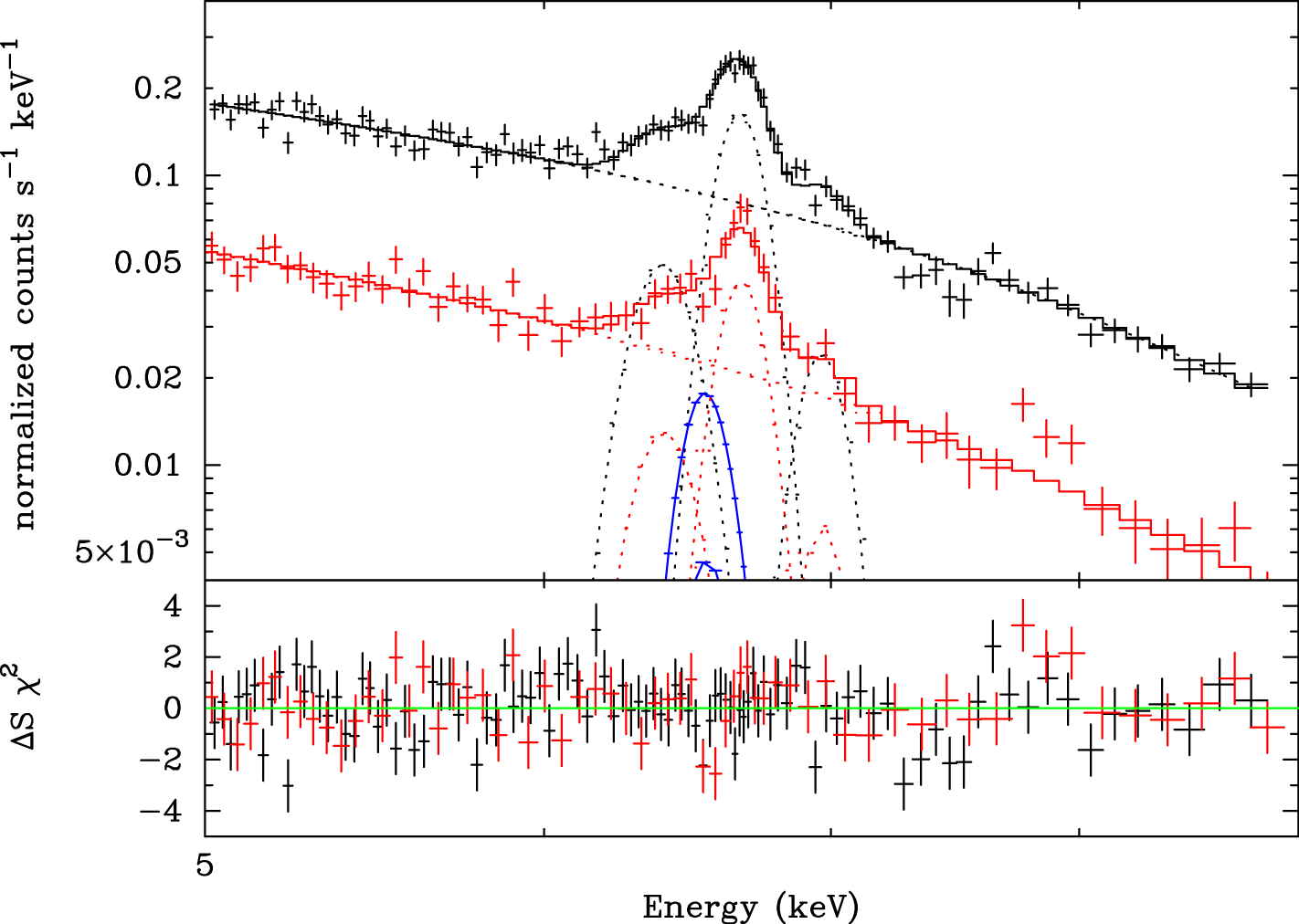}
 }
\caption{The outburst spectra in the 5--9~keV band with the best-fit
 model composed of a thermal bremsstrahlung and three iron lines
 including a Compton shoulder of the He-like {\ka} line in blue
 color. \label{fig:fit_comp_model}}
\end{figure}

We have attempted to evaluate the outburst spectra in the 5--9~keV band
again by implementing this Compton scattered He-like iron {\ka} line
component in the model. In the fitting, the central energies of the iron
lines are all fixed at their rest-frame energies (6.97~keV, 6.67~keV,
and 6.40~keV). The height of the plasma $h$ is not constrained very
well, and hence we also fix it at $0.1R_{\rm out}$, which is expected
from the best-fit value of $\Omega^{\rm O}/2\pi = 0.9$. The result of
the fit is shown in Fig.~\ref{fig:fit_comp_model}, and its best-fit
parameters are summarized in table~\ref{tab:fit_comp_model}.
\begin{table}[htb]
\caption{Best-fit parameters of a thermal bremsstrahlung plus three iron
 {\ka} lines including a Compton shoulder to the outburst spectra in the
 5--9~keV band. \label{tab:fit_comp_model}}
\begin{center}
\begin{tabular}{lll}
\hline\hline
Component & Parameter & Best-fit value \\
\hline
continuum & $kT$ (keV) & $8.0^{+0.6}_{-0.5}$ \\
 & normalization\footnotemark[$\dagger$] & $3.80^{+0.16}_{-0.14}$ \\
H-like line & energy (keV) & 6.97 (fixed) \\
 & normalization\footnotemark[\P] & $0.81^{+0.18}_{-0.17}$ \\
 & EW (eV) & $78^{+17}_{-16}$ \\
He-like line & energy (keV) & 6.67 (fixed) \\
 & normalization\footnotemark[\P] & $4.93^{+0.25}_{-0.30}$ \\
 & EW (eV) & $431^{+22}_{-26}$ \\
Compton & $h/R_{out}$ & 0.1 (fixed) \\
       & EW (eV) & $42^{+2}_{-3}$ \\
Neutral line & energy (keV)
       & 6.40 (fixed) \\
       & $\sigma$ (keV) & $0.073^{+0.079}_{-0.057}$ \\
       & normalization\footnotemark[\P] 
       & $1.74^{+0.67}_{-0.41}$ \\
       & EW (eV) & $140^{+54}_{-33}$ \\ \hline
 & $\chi^2$(d.o.f) & 205.7(157) \\
 & $\chi^2_{\nu}$ & 1.31 \\
\hline
\multicolumn{3}{@{}l@{}}{\hbox to 0pt{\parbox{85mm}{\footnotesize
\footnotemark[$\dagger$] In a unit of ($3.02\times10^{-18}/(4\pi D)^{2}\
   EM$), where $EM$ is the emission measure in a unit of ${\rm
   cm^{-3}}$. 
 \par\noindent
\footnotemark[$\ddagger$] In a unit of ($\times10^{-3}\ {\rm cm^{-2}\
   keV^{-1}\ s^{-1}}$) at 1~keV. 
 \par\noindent
\footnotemark[\P] In units of ($\times10^{-5}\ {\rm photons\ cm^{-2}\
   s^{-1}}$). 
}\hss}}
\end{tabular}
\end{center}
\end{table}
Here we have employed a thermal bremsstrahlung as the continuum because
it provides a better fit than a power law. The normalization of the
Compton shoulder is tied to that of He-like iron {\ka} line according to
the expectation from $h = 0.1R_{\rm out}$. As a result of the fit, we
found that the equivalent width of the broad 6.4~keV line is reduced
from $208^{+68}_{-27}$~eV to $140^{+54}_{-33}$~eV. This value is,
however, only marginally consistent with the equivalent width predicted
from the given reflection scale (${\rm
EW}(\Omega/2\pi\!=\!0.9^{+0.5}_{-0.4})\!=\!72^{+40}_{-32}$~eV).

The obtained Gaussian $\sigma$ of the 6.4~keV line, on the other hand, is
interpreted to represent the Keplerian velocity of the disk, which is
\begin{eqnarray}
\sqrt{\frac{GM_{WD}}{r}} \;=\; v_{\rm K}(r)
 \;=\; \left(\frac{\sigma}{E_0}\right)\ \frac{c}{\sin i},
\end{eqnarray}
where $E_0 = 6.4$~keV is the central energy of the fluorescent iron
{\ka} line, $i$ is the inclination angle $\simeq$37$^{\circ}$. From the
Gaussian $\sigma$ evaluated with the Compton shoulder model,
$\sigma\!=\!0.073^{+0.079}_{-0.057}$~keV
(table~\ref{tab:fit_comp_model}), we obtain $v_{\rm K}(r) =
5700^{+6100}_{-4400}$~km~s$^{-1}$, resulting in a radius of $r =
3.5^{+68.4}_{-2.4}\times 10^8$~cm for the 1.19$M_\odot$ white dwarf. The
optically thin thermal plasma seems to exist on the disk within $r
\lesssim 7\times 10^9$~cm from the white dwarf. It may be intriguing to
compare the spatial extension of the plasma thus obtained with that of
an eclipsing dwarf nova in outburst or a nova-like variable. Based on
the XMM-Newton observation of the nova-like variable UX~UMa,
\citet{2004MNRAS.348L..49P} measured an elapsed phase for the the hard
X-ray eclipse transition to be $\Delta \varphi \leq 0.01$.  The binary
parameters (\cite{2003A&A...404..301R} and references therein), on the
other hand, indicate an orbital separation of $a = 9.7\times
10^{10}$~cm. Hence, the plasma size is estimated to be $\leq 10^9$~cm,
in good agreement with the current estimation of SS~Cyg in outburst.

\subsection{Elemental Abundances}

\begin{figure}[htb]
\centerline{
 \includegraphics[width=0.49\textwidth,bb=0 0 991 746]{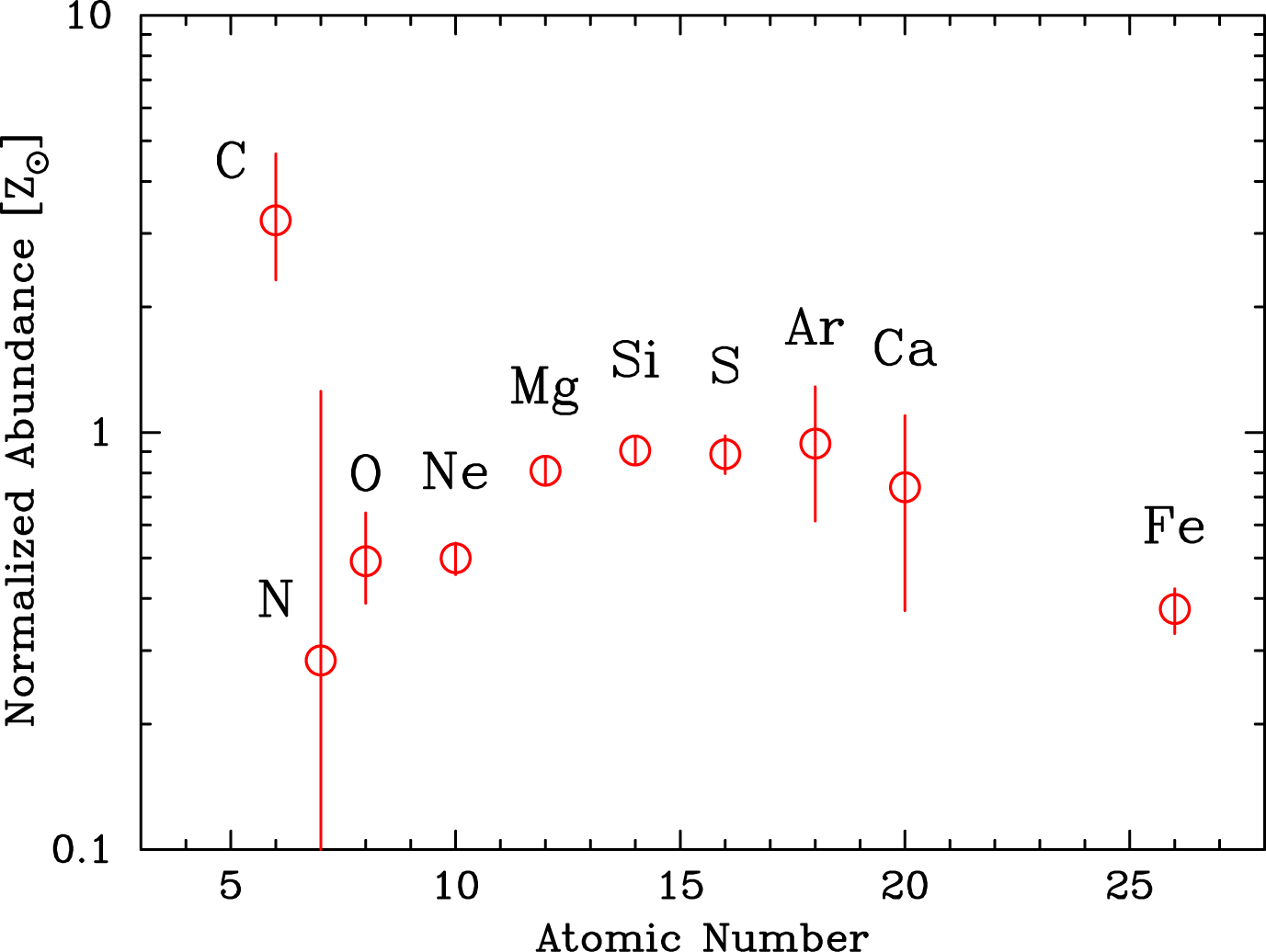}
 \includegraphics[width=0.49\textwidth,bb=0 0 991 746]{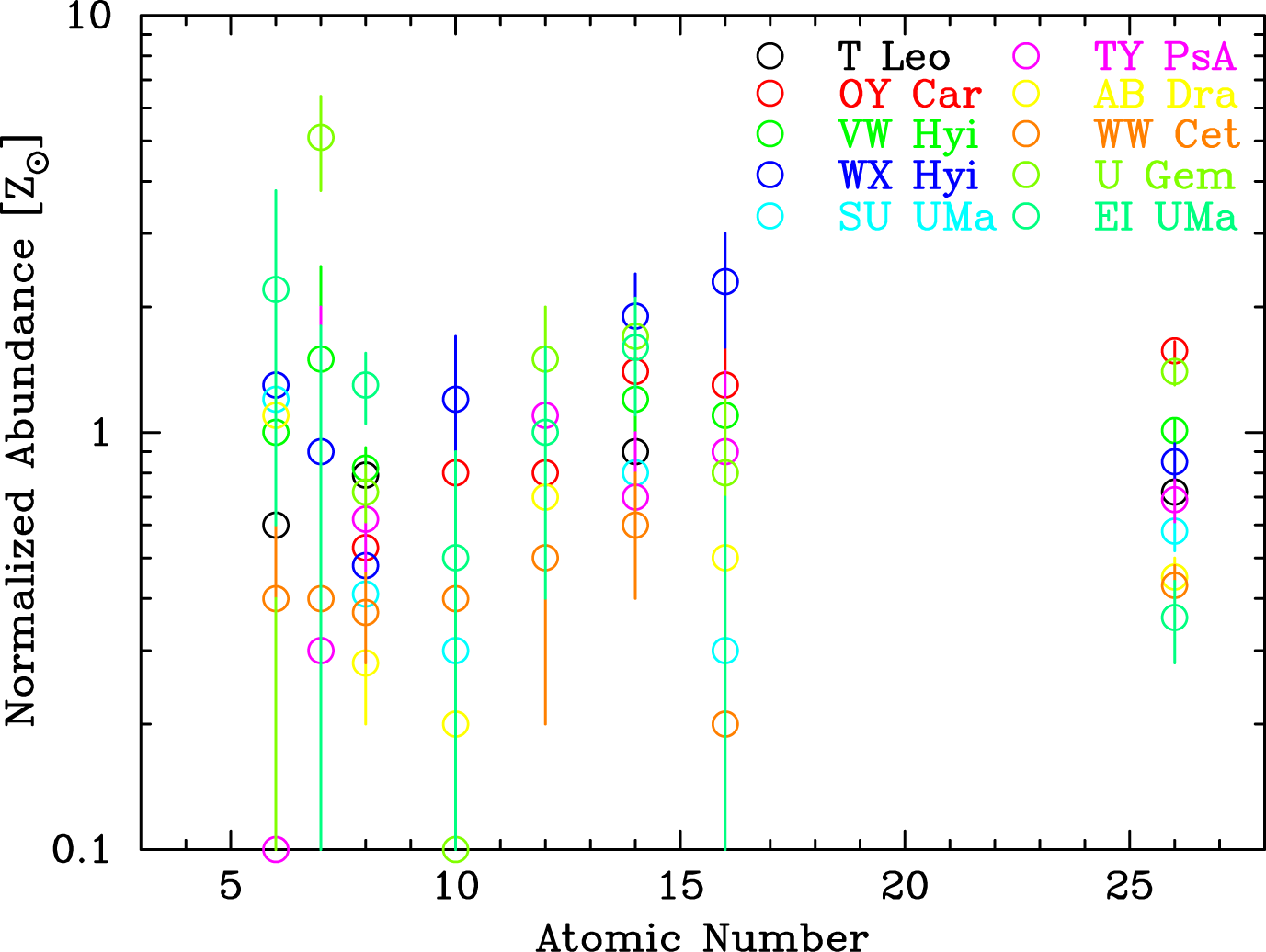}
}
\caption{(left) Abundances of SS~Cyg obtained with the Suzaku
 observations (\S~\ref{sec:abundances}). (right) Abundances of the 10
 non-magnetic CVs obtained with the XMM-Newton observations in
 quiescence \citep{2005ApJ...626..396P}. Abundances are given relative
 to the solar values in \citet{1989GeCoA..53..197A}
 \label{fig:abundance_summary}}
\end{figure}

The elemental abundances of SS~Cyg are summarized in
table~\ref{tab:abundances}, which are plotted in the left panel of
Fig.~\ref{fig:abundance_summary}.
All of the elemental abundances except for C are generally sub-solar.
\citet{1997MNRAS.288..649D} determined the abundance of Si, S, and Fe
separately, and that of the other elements in common by fitting ASCA
outburst spectrum, which are also subsolar. We further resolved the
abundances of Mg, Ne, O, N, and C. The right panel of
Fig.~\ref{fig:abundance_summary}, on the other hand, shows the
abundances of 10 non-magnetic CVs obtained with the XMM-Newton
observations \citep{2005ApJ...626..396P}. These abundances are
determined also by fits with the same model as ours. Their values are
widely distributed from sub-solar to near solar abundances. The
abundances of SS~Cyg are within these distribution in general.

%%%
\section{Conclusion}
%%%
We have presented results of the Suzaku observations on the dwarf nova
SS~Cyg in quiescence and outburst in 2005 November. The X-ray spectra of
SS~Cyg are composed of a multi-temperature optically thin thermal plasma
model with a maximum temperature of a few tens of keV, its reflection
from the white dwarf surface and/or the accretion disk, and a 6.4~keV
neutral iron {\ka} line from the reflectors via fluorescence. High
sensitivity of the HXD PIN detector and the high spectral resolution of
the XIS enable us to disentangle degeneracy between the maximum
temperature and the reflection parameters, and to determine the emission
parameters with unprecedented precision. The maximum temperature of the
plasma in quiescence $kT^{\rm Q}_{\rm max} =
20.4^{+4.0}_{-2.6}\,\mbox{(stat.)}\pm 3.0\,\mbox{(sys.)}$~keV is
significantly higher than that in outburst $kT^{\rm O}_{\rm max} =
6.0^{+0.2}_{-1.3}$~keV. The elemental abundances of the plasma are close
to the solar ones for the medium-Z elements (Si, S, Ar) whereas they
declines both in lighter and heavier elements. Those of oxygen and iron
are 0.46$^{+0.04}_{-0.03}\,\mbox{(stat.)}\pm 0.01\,\mbox{(sys.)}Z_\odot$
and 0.37$^{+0.01}_{-0.03}\,\mbox{(stat.)}\pm
0.01\,\mbox{(sys.)}Z_\odot$. The exception is carbon whose abundance is
at least $2Z_\odot$ even if we take into account all possible systematic
errors. These trends are similar to other dwarf novae observed with
XMM-Newton \citep{2005ApJ...626..396P}.

The solid angle of the reflector subtending over the optically thin
thermal plasma is $\Omega^{\rm Q}/2\pi = 1.7\pm 0.2\,\mbox{(stat.)}\pm
0.1\,\mbox{(sys.)}$ in quiescence. Since even an infinite slab can
subtend a solid angle of $\Omega/2\pi = 1$ over a radiation source above
it, this large solid angle can be achieved only if the plasma views both
the white dwarf and the accretion disk with substantial solid
angles. Thanks to high energy resolution of the XIS, we have resolved a
6.4~keV iron {\ka} line into a narrow and broad components (significance
of the broad component is $\sim$99\%), which also indicate contributions
from both the white dwarf and the accretion disk to the reflected
continuum spectra. The equivalent widths of them are both
$\sim$50~eV. From all these results, we consider the standard optically
thin BL formed between the inner edge of the accretion disk
and the white dwarf surface \citep{1985ApJ...292..535P} as the most
plausible model to explain the observed large solid angle. From the
equivalent width of the narrow 6.4~keV component, the height of the
BL from the white dwarf surface is $h < 0.12R_{\rm WD}$. The
total equivalent width of the 6.4~keV line ($\sim$100~eV) is consistent
with that expected from $\Omega^{\rm Q}/2\pi$, the iron abundance, and
the incident illuminating continuum spectrum.

The solid angle of the reflector in outburst $\Omega^{\rm O}/2\pi =
0.9^{+0.5}_{-0.4}$, on the other hand, is significantly smaller than
that in quiescence, and is consistent with an infinite slab. Since the
6.4~keV iron emission line is broad with no narrow component
($\lesssim$20\% of the broad component), the reflection originates from
the accretion disk. The accretion belt can also contribute to the
reflection. The 6.4~keV line from the accretion belt is expected to be
broad, which is consistent with the absence of the narrow 6.4~keV
component. The EW of the 6.4~keV line is so large that it cannot be
interpreted within a simple scheme of reflection from the disk. Even if
Compton down-scattering of the observed He-like {\ka} line is taken into
account, we can only find a solution which marginally reconciles the
large EW with the solid angle of the reflector. We consider the
optically thin thermal plasma in outburst as being distributed on the
accretion disk. The Chandra HETG observation in outburst revealed that
the He-like and H-like emission lines from O, Ne, Mg, and Si are broad
and their widths ($\sim$2000~km~s$^{-1}$) are consistent with those
expected from the Keplerian velocity of the accretion disk
\citep{2008ApJ...680..695O}. This fact suggests that the optically thin
thermal plasma is anchored to the accretion disk and the accretion belt
by magnetic field, for example, like solar coronae.

\appendix

\section{Low energy response of XIS1 \label{appendix:A}}

In order to analyze spectra with XIS1 in the energy band
$\lesssim$0.5~keV where the energy response is affected by accumulation
of contaminating material, we have checked the energy response of the
XIS-1 with the PKS2155--304 data taken between 2005 November 30 and 2005
December 1 (seq.\#700012010), which were taken close enough to our
SS~Cyg observations. Although the observation is originally planned to
continue for a 60~ks effective exposure time, part of the observation
suffered a light-leak accident and the CCD chip is irradiated by optical
and UV
photons\footnote{http://wwwxray.ess.sci.osaka-u.ac.jp/~hayasida/ftp-files/XIS2/XIS\_20060710b.pdf}.
After removing these time intervals, 39.4~ks data remain in total.  We
then created the same on-source and background regions as those adopted
in the SS~Cyg observations, and created the response file of XIS1. The
default correction was made for the contamination.

\begin{figure}[htb]
\centerline{
 \includegraphics[width=0.49\textwidth,bb=0 0 973 770]{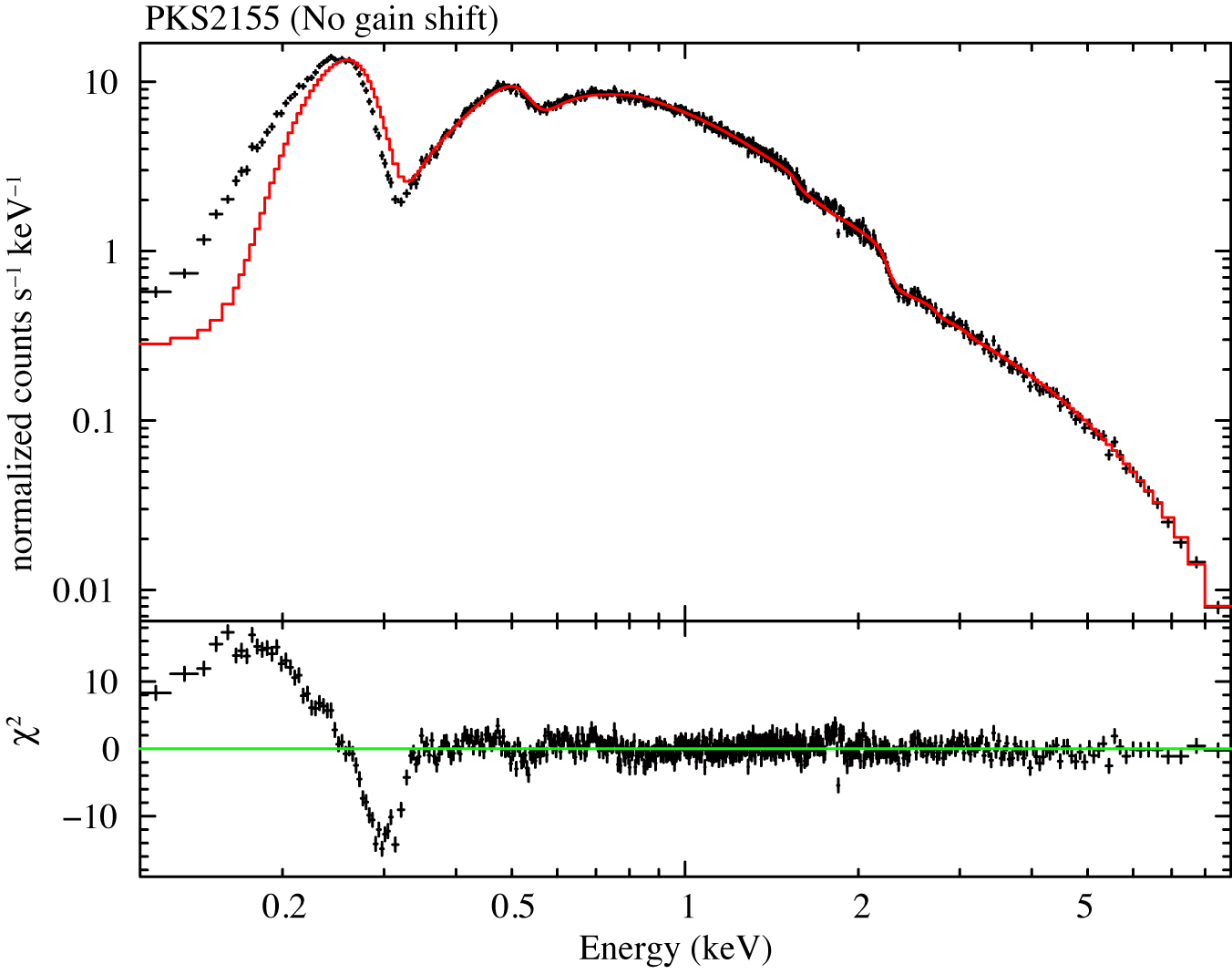}
 \includegraphics[width=0.49\textwidth,bb=0 0 976 770]{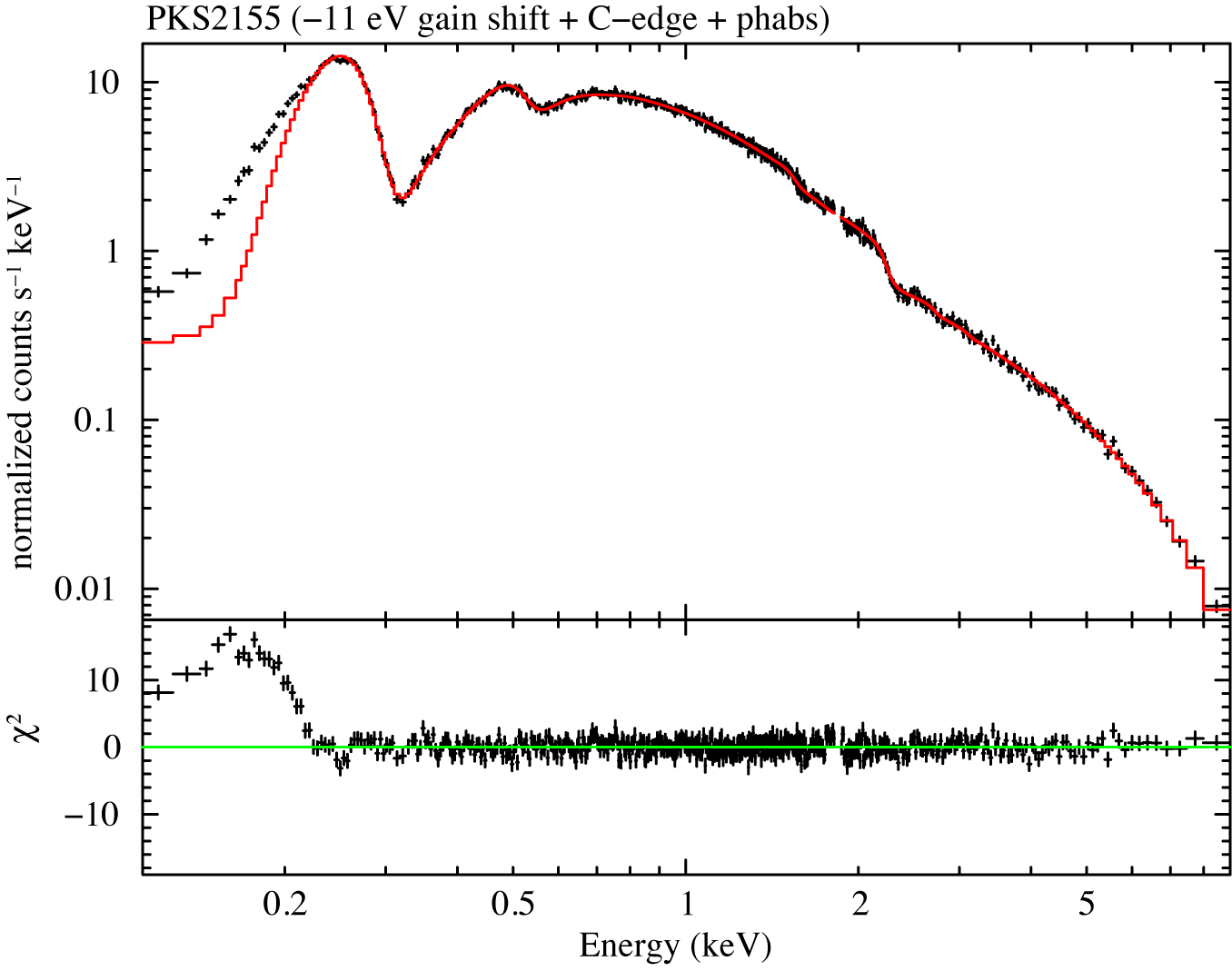}
}
\caption{Spectral fitting of XIS-1 in the energy range of
 0.1--10~keV before adjusting the energy scale (left), and after
 adjusting the gain and applying additional absorption. After these
 treatments, the energy band above 0.226~keV can safely be used for
 spectral analysis. \label{fig:appA}}
\end{figure}

Between 0.1 and 10~keV, PKS2155$-$304 has been well represented by a
curved spectrum with an energy slope gradually steepening from 1.1 to
1.6 \citep{1998A&A...333L...5G}. We therefore fit our XIS1 spectrum with
a broken power-law model
\begin{eqnarray}
F(E) &=& K(E/1~{\rm keV})^{-\Gamma_1}\qquad\qquad\quad (for~~ E\le E_c) \\
     &=& KE_c^{\Gamma_2 - \Gamma_1}\times (E/1~{\rm keV})^{-\Gamma_2}
     ~~(for~~ E\ge E_c),
\end{eqnarray}
where $\Gamma_1$ and $\Gamma_2$ are power law photon indexes, $K$ is
normalization factor, $E_c$ is a break energy, and see if the low energy
calibration is enough for our analysis. The hydrogen column density to
PKS2155 is $16.9\times10^{19}~{{\rm cm^{-2}}}$ from past infrared
observations. The result of fit is shown in the left panel of
Fig.~\ref{fig:appA}.
The fit below $\sim$0.35~keV is not very well, and there is evidence of
gain shift between 0.5--0.6~keV. Hence we apply an energy offset to the
model by 1~eV step, and found that the $-$11~eV shift gives the smallest
reduced $\chi^2$ value. The remaining residuals are adjusted by
introducing additional hydrogen column density and a carbon K edge at
0.2842~keV, which are both deemed associated with the contaminant of the
XIS. The result of the fit is shown in the right panel of
Fig.~\ref{fig:appA} and its best-fit parameters are summarized in
table~\ref{tab:appA}.
\begin{table}[ht]
\begin{center}
\caption{Best-fit parameters of additional absorption component to fit
 the spectrum of PKS2155$-$304.}
\label{tab:appA}
\begin{tabular}{ll} \hline \hline
Model & parameter \\ \hline
Gain offset (eV) & $-$11 (fixed)\\
$\tau$ of carbon K edge & $0.88\pm 0.05$ \\
$N_{\rm H}$ ($10^{19}$ cm$^{-2}$) & $8.2\pm 0.7$ \\ \hline
$\chi^2$ (d.o.f.) & 640 (547) \\ \hline
\end{tabular}
\end{center}
\end{table}
The additional photoelectric absorption and the edge significantly
improves the fit in 0.3--0.5~keV, and the spectrum down to 0.226~keV can
be well represented by the broken power-law model.  In the analysis of
the Suzaku SS~Cyg data, we always apply additional absorption column
density of $N_{\rm H} = 8.2\times10^{19}$~cm$^{-2}$ and the carbon edge
with an optical depth of $\tau = 0.88$ at 0.2842~keV. The amount of
gain offset is different from observation to observation. We have
checked the offsets during the Suzaku observations of SS~Cyg in
quiescence and outburst in the same way as is described above to find
that no offset ($\lesssim 2$~eV) is required for both observations.

\end{document}